\documentclass{emulateapj}
\shorttitle{Extended CO around VY~CMa and Betelgeuse}
\shortauthors{Smith, Hinkle, \& Ryde}
\begin{document}

\title{Red supergiants as potential Type II\lowercase{n} supernova
  progenitors: Spatially resolved 4.6~$\micron$ CO emission around
  VY~CM\lowercase{a} and Betelgeuse\altaffilmark{1}}

\author{Nathan Smith}
\affil{Astronomy Department, University of California, 601 Campbell
  Hall, Berkeley, CA 94720; nathans@astro.berkeley.edu}

\author{Kenneth H.\ Hinkle}
\affil{National Optical Astronomy Observatory, 950 North Cherry Ave.,
  Tucson, AZ 85719; hinkle@noao.edu}

\and

\author{Nils Ryde}
\affil{Lund Observatory, Box 43, SE-221 00, Lund, Sweden;
  ryde@astro.lu.se}

\altaffiltext{1}{Based on observations obtained at the Gemini
  Observatory.}

\begin{abstract}

  We present high-resolution 4.6~$\micron$ CO spectra of the
  circumstellar environments of two red supergiants (RSGs) that are
  potential supernova (SN) progenitors: Betelgeuse and VY Canis
  Majoris.  Around Betelgeuse, $^{12}$CO emission within
  $\pm$3\arcsec\ ($\pm$12 km~s$^{-1}$) follows a mildly clumpy but
  otherwise spherical shell, smaller than its $\sim$55\arcsec\ shell
  in K~{\sc i} $\lambda$7699. In stark contrast, 4.6~$\micron$ CO
  emission around VY~CMa is coincident with bright K~{\sc i} in its
  clumpy asymmetric reflection nebula, within $\pm$5\arcsec\ ($\pm$40
  km~s$^{-1}$) of the star.  Our CO data reveal redshifted features
  not seen in K~{\sc i} spectra of VY~CMa, indicating a more isotropic
  distribution of gas punctuated by randomly distributed asymmetric
  clumps.  The relative CO and K~{\sc i} distribution in Betelgeuse
  arises from ionization effects within a steady wind, whereas in
  VY~CMa, K~{\sc i} is emitted from skins of CO cloudlets resulting
  from episodic mass ejections 500--1000 yr ago.  In both cases, CO
  and K~{\sc i} trace potential pre-SN circumstellar matter: we
  conclude that an extreme RSG like VY~CMa might produce a Type~IIn
  event like SN~1988Z if it were to explode in its current state, but
  Betelgeuse will not.  VY~CMa demonstrates that luminous blue
  variables (LBVs) are not necessarily the {\it only} progenitors of
  SNe~IIn, but it underscores the requirement that SNe~IIn suffer
  enhanced episodic mass loss shortly before exploding.

\end{abstract}

\keywords{circumstellar matter --- stars: evolution --- stars: mass
  loss --- stars: winds, outflows}

\section{INTRODUCTION}
				   
Massive stars shed copious mass before exploding as supernovae (SNe)
with a wide diversity of observable properties.  Stars in the initial
mass range 20--40~$M_{\odot}$ are key, because their complex
post-main-sequence evolution is dictated to a large extent by their
poorly understood mass loss.  They represent a transitional range of
mass below which stars die in a normal red supergiant (RSG) phase, and
above which they skip the RSG phase altogether and become luminous
blue variables (LBVs) instead.  In between, stars may traverse the HR
Diagram through a cool RSG phase with strong mass loss, but they can
potentially evolve back to warmer temperatures through a variety of
hypergiant phases to a Wolf-Rayet (WR) phase if mass loss is
sufficient to remove the H envelope.  Depending on when they explode
along this mass-loss sequence, stars in this initial mass range can
appear as SNe of Types II-P, II-L, IIb, or Ib/c -- or if the immediate
pre-SN mass-loss is severe, they could potentially be Type~IIn with
relatively narrow H emission lines (see Filippenko 1997 for a review
of SN types).  Here we investigate under which conditions this is
likely.

The amount and specific nature of this RSG mass loss can dictate the
subsequent evolution of the star, especially its chemical mixing, mass
stripping, and angular momentum loss.  At the present time, however,
the driving of RSG winds is poorly understood and mass-loss rates are
not calculable from {\it ab initio} theory, so stellar evolution
models must adopt mass-loss rates guided by observations. These are
usually time-averaged mass-loss rates, but $\dot{M}(t)$ may vary
prodigiously during RSG evolution.  Empirical studies are critical to
constrain physical parameters of the mass loss, in order to decipher
how observed SN types map to initial mass.  This may be critical for
phases just before explosion.

Like their more massive siblings the classical LBVs (Smith \& Owocki
2006; Humphreys \& Davidson 1994), stars of initial mass 20--40
$M_{\odot}$ can display relatively sudden, episodic bursts of mass
loss in their post-main-sequence evolution.  This can happen as RSGs,
as yellow hypergiants (de Jager 1998; de Jager \& Nieuwenhuijzen
1997), or in a post-RSG LBV phase (Smith et al.\ 2004).  As is the
case with LBV eruptions, this episodic mass loss is not accounted for
in current models of stellar evolution, but it may nevertheless
significantly alter a star's evolution (Smith \& Owocki 2006).
Besides LBVs, examples of episodic mass loss are the outbursts of
variable yellow hypergiants like $\rho$~Cas and Var~A in M~33 (Lobel
et al.\ 2003; Humphreys et al.\ 2006; de Jager 1998; de Jager \&
Nieuwenhuijzen 1997), and the nebulae of stars like VY~CMa (Smith et
al.\ 2001; Smith 2004; Decin et al.\ 2006; Humphreys et al.\ 2005,
2007).  If the dense circumstellar material (CSM) from these outbursts
still surrounds the star when it explodes, it would provide a dense
obstacle for the SN blast wave to overtake.

This brings us back to the likely progenitors of Type~IIn supernovae.
Judging by their dense CSM resulting from episodic mass ejection, LBVs
are likely to appear as SNe~IIn if they explode within $\la$10$^3$ yr
after a major outburst (Smith et al.\ 2007, 2008; Gal-Yam et al.\
2007).  This defies current stellar evolution models, however, which
predict that at solar metallicity, stars massive enough to make LBVs
should end their lives as WR stars after shedding their H envelopes
(e.g., Heger et al.\ 2003; Meynet et al.\ 1994; Maeder \& Meynet 1994;
Woosley et al.\ 1993).  Possible exceptions are the lower-luminosity
LBVs that are post-RSGs (Smith et al.\ 2004).

Because mass-loss rates in steady line-driven winds are inadequate to
shed the star's H envelope, Smith \& Owocki (2006) have argued that
the mass deficit must be shed in LBV eruptions, or if not, that the
star will explode as an LBV with much of its H envelope intact (see
also Smith 2007; Smith et al.\ 2007, 2008; Gal-Yam et al.\ 2007). As
emphasized elsewhere (Smith 2008; Smith et al.\ 2007), continually
mounting observational evidence suggests that LBVs or LBV-like stars
may explode (Smith \& Owocki 2006; Smith et al.\ 2007, 2008; Gal-Yam
et al.\ 2007; Vink et al.\ 2008; Kotak \& Vink 2006).  The recent Type
IIn events SN~2006gy and 2006tf were spectacular examples of this
phenomenon (Smith et al.\ 2007, 2008).  Gal-Yam et al.\ (2007)
identified an LBV-like progenitor of a SN~IIn, and postulated that
essentially all SNe~IIn mark the explosions of LBVs.

There are other less contentious candidates for SNe~IIn progenitors,
however, which remain to be considered in detail. The essential
requirement for a Type~IIn progenitor, as emphasized by Smith et al.\
(2008), is extreme mass-loss with H-rich material shortly before
explosion.  The most likely alternatives to LBVs, all with dense CSM,
are: the most extreme RSGs like VY Canis Majoris (Smith et al.\ 2001),
yellow hypergiants like IRC+10420 (Humphreys et al.\ 1997, 2002;
Davies et al.\ 2007), or B[e] supergiants (Zickgraf et al. 1996).

In this paper, we examine the dense circumstellar environment of
VY~CMa as it compares to that of a more standard RSG like Betelgeuse.
We have chosen to use the low-excitation fundamental
vibration-rotation band-head lines of the $R$-branch of $^{12}$CO at
$\sim$4.6~$\micron$ as a tracer of the densest gas in the CSM.  We
selected three lines: $^{12}$CO 1--0 $R$1 46493.11 \AA, 1--0 $R$2 at
46412.42 \AA, and 1--0 $R$3 at 46332.76 \AA (vacuum wavelengths).
These lines were first detected in the spectrum of Betelgeuse by
Bernat et al.\ (1979), who modeled them as scattered lines, and the
detection of related CO lines in VY~CMa was reported by Geballe et
al.\ (1973).  In both cases, CO was seen in {\it absorption} in the
spectrum of the central star.  Following a preliminary study of the
4.6~$\micron$ CO lines in emission in Betelgeuse (Ryde et al.\ 1999),
we report a detailed study of spatially extended emission in the IR
vibration-rotation lines around these stars.  Spatially extended
emission probes the CSM kinematics directly.

Both Betelgeuse and VY~CMa also exhibit spatially resolved K~{\sc i}
$\lambda$7699 emission in their nebulae, the details of which will be
reviewed later.  The CO lines we investigate here have several
advantages over K~{\sc i} and other probes of the CSM.  The
interpretation of scattered atomic resonance lines like K~{\sc i} and
Na~{\sc i} is complicated by subtle changes in the ionization state,
atoms can be depleted by grain formation to an uncertain degree, and
at visual wavelengths these lines can be strongly affected by
extinction within dusty envelopes.  Observing CO at longer infrared
(IR) wavelengths mitigates these difficulties.  Also, long-slit
spectra at IR wavelengths with array detectors allow higher spatial
resolution than is usually applied in molecular observations at mm
wavelengths (e.g., Muller et al.\ 2007).

We present our new CO observations in \S 2.  In \S3 and \S4, for
Betelgeuse and VY~CMa respectively, we give some background of related
previous observational work, and we discuss how the observed CO
structure samples their CSM environments -- particularly with regard
to the K~{\sc i} emission.  In \S 5 we summarize our results and
speculate about the consequences if these objects explode as
core-collapse SNe in their current states.

%%%%%%%%%%%%%%%%%%%%%%%%% FIGURE 1 - Image A Ori %%%%%%%%%%%%%%%%%%%%%
\begin{figure}
\epsscale{1.0}
\plotone{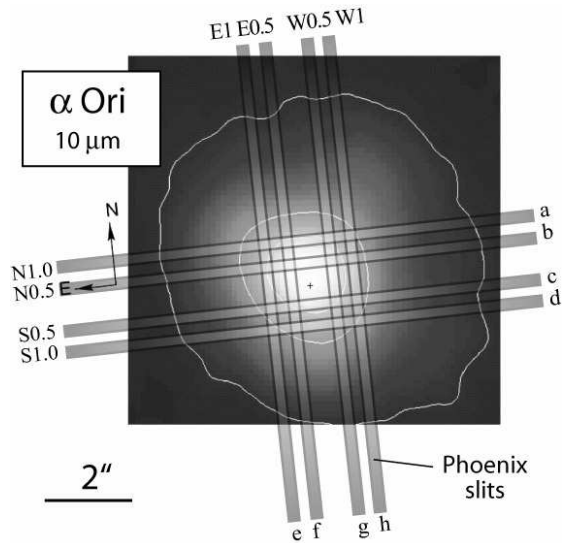}
\caption{A mid-IR image of Betelgeuse with the placements of our 8
  individual Phoenix slit positions marked.  The image is a
  10~$\micron$ nulling interferometer image from Hinz et al.\ (1998;
  with permission), where north is rotated slightly left of vertical
  in their original image.  Because of the technique used, this image
  emphasizes extended nebular dust emission, rather than photospheric
  emission.  The lower-case letters for each slit position correspond
  to the individual panels in Figure~4.}
\label{fig:imgOri}
\end{figure}
%%%%%%%%%%%%%%%%%%%%%%%%%%%%%%%%%%%%%%%%%%%%%%%%%%%%%%%%%%%%%%%%%%%%%%

%%%% TABLE 2 - SPECTRA
\begin{deluxetable*}{llccl}\tabletypesize{\scriptsize}
\tablecaption{Phoenix Observation Log}
\tablewidth{0pt}
\tablehead{
  \colhead{Date} &\colhead{Target} &\colhead{P.A.}
  &\colhead{Exp.\ Time} &\colhead{Offsets} 
}
\startdata
2005 Feb 28 &Betelgeuse   &0\arcdeg    &30 s   &star (Lyot Hartman mask) \\
2005 Feb 28 &Betelgeuse   &0\arcdeg    &120 s  &1$\farcs$0 E, 0$\farcs$5 E, 0$\farcs$5 W, 1$\farcs$0 W \\
2005 Feb 28 &Betelgeuse   &90\arcdeg   &120 s  &1$\farcs$0 N, 0$\farcs$5 N, 0$\farcs$5 S, 1$\farcs$0 S \\
2006 Dec 1  &VY CMa   &0\arcdeg &4 s &star \\
2006 Dec 1  &VY CMa   &0\arcdeg &120 s &2$\farcs$0 E, 1$\farcs$5 E,1$\farcs$0 E, 0$\farcs$5 E, 0$\farcs$5 W, 1$\farcs$0 W, 1$\farcs$5 W, 2$\farcs$0 W, 2$\farcs$5 W \\
2006 Dec 1  &VY CMa  &90\arcdeg &120 s &2$\farcs$0 N, 1$\farcs$5 N,1$\farcs$0 N, 0$\farcs$5 N, 0$\farcs$5 S, 1$\farcs$0 S, 1$\farcs$5 S, 2$\farcs$0 S \\
\enddata
%\tablenotetext{a}{}
%\tablecomments{.......}
\end{deluxetable*}

%%%%%%%%%%%%%%%%%%%%%%%%% FIGURE 2 - Images %%%%%%%%%%%%%%%%%%%%%%%%%%%
\begin{figure*}
\epsscale{1.0}
\plotone{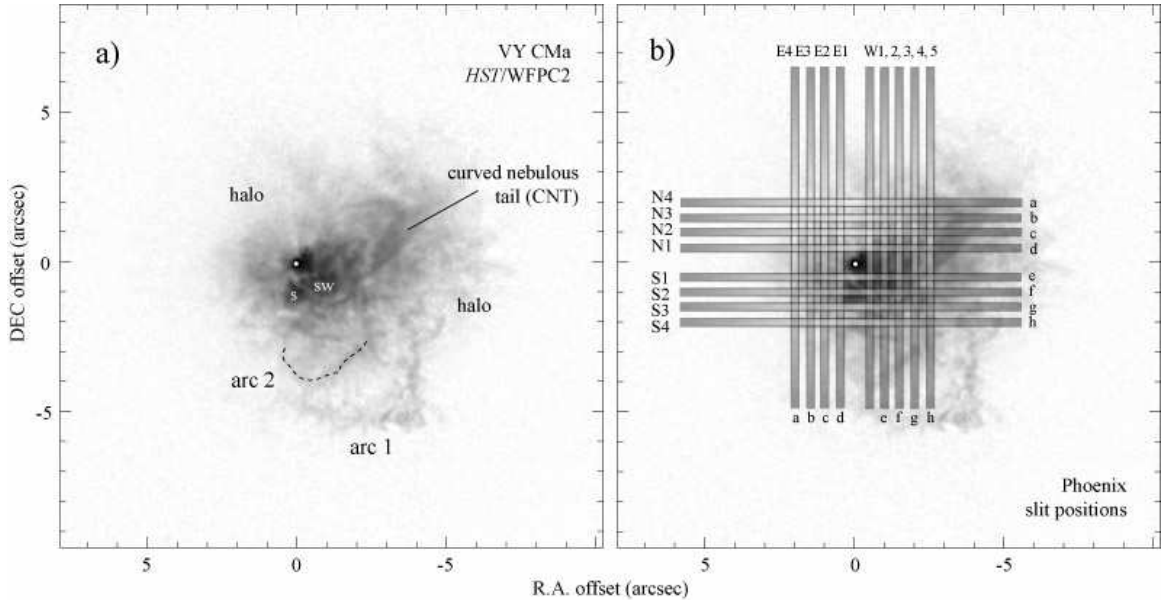}
\caption{ (a) Visual-wavelength image of VY~CMa made with {\it
    HST}/WFPC2 (from Smith et al.\ 2001) with the same labels for
  features in the nebula given by Smith (2004).  (b) Phoenix slit
  aperture positions superposed on the same image of the nebula.  Slit
  placements are in increments of 0$\farcs$5, beginning at 0$\farcs$5
  from the star in all four directions.  The letters for each slit
  position correspond to the individual panels in
  Figs.~\ref{fig:velVYew} and \ref{fig:velVYns} (W1, positioned
  0$\farcs$5 west of the star, is not shown in any panel in
  Fig.~\ref{fig:velVYew} because of saturation).}
\label{fig:img}
\end{figure*}
%%%%%%%%%%%%%%%%%%%%%%%%%%%%%%%%%%%%%%%%%%%%%%%%%%%%%%%%%%%%%%%%%%%%%%

%%%%%%%%%%%%%%%%%%%%%%%%% FIGURE 3 - A Ori spectrum %%%%%%%%%%%%%%%%%
\begin{figure*}[!ht]
\epsscale{0.92}
\plotone{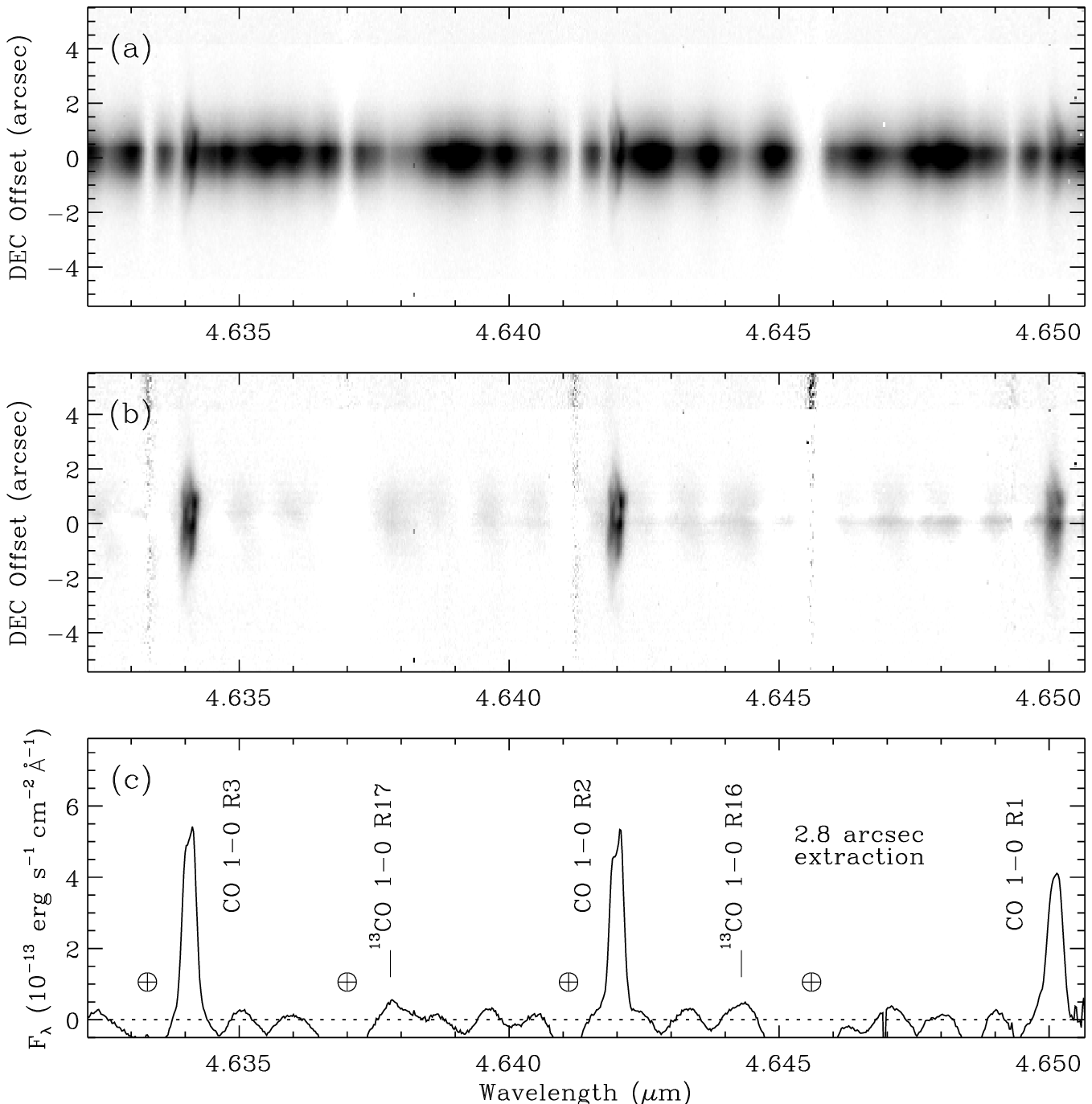}
\caption{Phoenix spectra of Betelgeuse at one example slit position
  1\arcsec\ west of the star, demonstrating the typical appearance of
  the spectra over the full wavelength range.  The top panel (a) shows
  the original 2-D long-slit spectrum.  The middle panel (b) is the
  same as the top, but with the scattered starlight subtracted out by
  scaling the central star's spectrum to the spatial profile of
  scattered light at each pixel along the slit, leaving only intrinsic
  CO emission.  The bottom panel (c) is a 1-D extraction of the bright
  inner part of the 2-D star-subtracted spectrum in Panel b using a
  2$\farcs$8$\times$0$\farcs$17 extraction aperture.  Places where
  strong telluric features left severe subtraction residuals are
  marked.  Possible detections of $^{13}$CO 1--0 R17 and R16 are also
  marked.}
\label{fig:fullORI}
\end{figure*}
%%%%%%%%%%%%%%%%%%%%%%%%%%%%%%%%%%%%%%%%%%%%%%%%%%%%%%%%%%%%%%%%%%%%%%

\section{OBSERVATIONS}

High-resolution ($R\simeq$75,000; $\sim$4 km s$^{-1}$) IR spectra of
$\alpha$~Ori and VY~CMa were obtained on 2005 Feb.\ 27 \& 28, and 2006
Dec.\ 1, respectively.  The spectral range was centered on the
4.6~$\micron$ fundamental vibration-rotation bandhead emission of CO,
using the Phoenix spectrograph (Hinkle et al.\ 1998) on the Gemini
South telescope.  Phoenix has a 256$\times$1024 InSb detector with a
pixel scale of 0$\farcs$085 $\times$ 1.8 km s$^{-1}$ at a wavelength
of $\sim$4.6~$\micron$.  Sky conditions were photometric, and the
near-IR seeing was typically 0$\farcs$5.  Removal of airglow lines and
thermal sky emission was accomplished by subtracting an observation of
an off-source sky position far from the star immediately after each
set of targeted observations.

HR~1713 ($\beta$~Ori) and HR~2618 were observed with Phoenix on the
same nights with the same grating settings in order to correct for
telluric absorption and for flux calibration of the spectra of
Betelgeuse and VY~CMa, respectively.  Absolute flux-calibration
uncertainties are dominated by the alignment of the standard stars in
the slit, and may be as high as $\pm$25\%, although relative
uncertainties in the flux from one slit position to the next are much
less.  Telluric lines were used for wavelength calibration, using the
telluric spectrum available from Hinkle et al.\ (1995).  For both
targets, intrinsic CO emission is redshifted out of the telluric CO
absorption features.  Velocities were calculated adopting a vacuum
rest wavelength of 46412.42 \AA \ for $^{12}$CO 1--0 R2.  These
velocities were then corrected to a heliocentric reference frame, with
the uncertainty of $\pm$1 km s$^{-1}$ dominated by scatter in the
dispersion solution for telluric lines.

We used a 0$\farcs$17-wide (2-pixel) long-slit aperture, except for
the on-star spectrum of Betelgeuse for which we employed a Lyot
Hartman mask to avoid saturation.  For each source, the Phoenix slit
was oriented at two different perpendicular position angles running
N-S and E-W.  To sample the kinematics across each nebula, the slit
was positioned on the bright central star, plus several offsets
stepped in increments of 0$\farcs$5 in either direction perpendicular
to the slit axis (Table 1).  In order to show correspondence with
features in the nebulosity, Figure~\ref{fig:imgOri} shows slit
positions on a mid-IR image of Betelgeuse (Hinz et al.\ 1998), and
Figure~\ref{fig:img} shows slit positions on a visual-wavelength image
of VY~CMa (Smith et al.\ 2001) obtained with the {\it Hubble Space
  Telescope} ({\it HST}).

\begin{figure*}
\epsscale{0.92}
\plotone{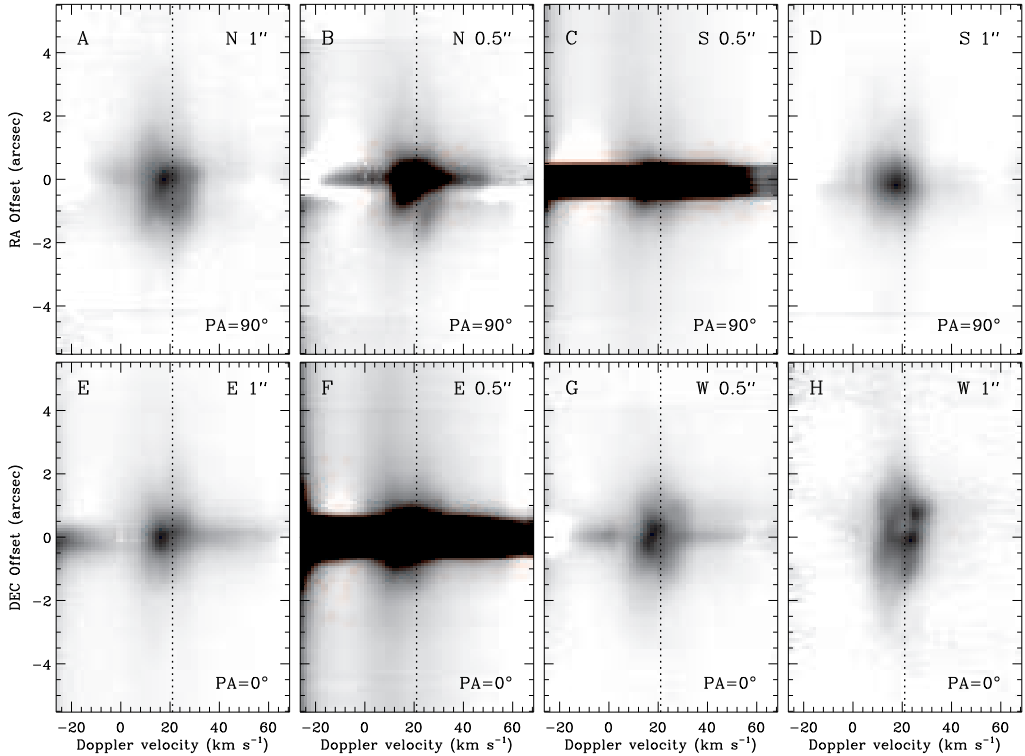}
\caption{Spatial-kinematic structure of extended CO 1--0 R2 emission
  around Betelgeuse at several different Phoenix slit positions, taken
  from a section of star-subtracted spectra as in
  Figure~\ref{fig:fullORI}$b$.  Slit position angles and offset
  positions from the star are indicated in each panel.  Velocities are
  plotted as the heliocentric Doppler velocity, and the systemic
  velocity of +21 km s$^{-1}$ is marked.}
\label{fig:velORI}
\end{figure*}

Because we aim to detect faint line emission near a bright star, our
data reduction employed additional steps beyond standard spectral
reduction in order to enhance the visibility of extended CO emission.
After standard long-slit spectral reduction, we had a series of
flux-calibrated 2-D spectra.  An example of a single offset position
for Betelgeuse is shown in Figure~\ref{fig:fullORI}a.  This spectrum
is dominated by the bright emission from the central star, including
instrumental and atmospheric scattering of the direct stellar light,
plus starlight scattered by circumstellar dust.  CO emission lines can
be seen in this spectrum, but are overwhelmed by starlight and are
difficult to distinguish amid the complex spectrum of the star.  We
subtracted out the direct and scattered photospheric light with a
model for each individual slit position by taking the directly
observed 1-D spectrum of the central star, wherein circumstellar CO
emission is negligible compared to the bright photosphere, and scaling
it along each spatial pixel with a spatial sample of the intensity in
a region of the spectrum free from CO bandhead emission.  The result
of this subtraction is shown in Figure~\ref{fig:fullORI}b, and a 1-D
tracing of the star-subtracted spectrum is shown in
Figure~\ref{fig:fullORI}c.  Figure~\ref{fig:velORI} shows
position-velocity plots of the star-subtracted 1--0 R2 line for all
observed slit positions.  Some positions close to the star were
affected by saturation.  We also show spectral and spatial tracings in
Figures~\ref{fig:traceORI} and \ref{fig:radORI}, respectively, and
these will be discussed later.

We performed the same type of scattered-light subtraction procedure
for our VY~CMa data, even though the CO emission is clearly seen even
in the raw spectra.  Examples of the 2-D spectra before and after
subtracting scattered starlight are shown in Figures~\ref{fig:fullVY}a
and \ref{fig:fullVY}b, respectively, and an example 1-D tracing is
shown in Figure~\ref{fig:fullVY}c.  As for Betelgeuse, we produced a
series of 2-D position-velocity plots of the star-subtracted 1--0 R2
line of CO for all slit offset positions.  These are displayed in
Figure~\ref{fig:velVYew} for the observations at P.A.=0\arcdeg\ offset
to the E and W of the star, and in Figure~\ref{fig:velVYns} for
P.A.=90\arcdeg\ offset N and S of the star's position.  Spectral
tracings of selected positions are plotted in Figure~\ref{fig:traceVY}
and radial spatial intensity tracings are plotted in
Figure~\ref{fig:radVY}.

%%%%%%%%%%%%%%%%%%%%%%%%% FIGURE 4 - A Ori stamps %%%%%%%%%%%%%%%%%%%%
%%%%%%%%%%%%%%%%%%%%%%%%%%%%%%%%%%%%%%%%%%%%%%%%%%%%%%%%%%%%%%%%%%%%%%

%%%%%%%%%%%%%%%%%%%%%%%%%%%%%%%%%%%%%%%%%%%%
%%%%%%%%%%%%%%%%%%%%%%%%%%%%%%%%%%%%%%%%%%%%
\section{BEETLEJUICE, BEETLEJUICE, BEETLEJUICE}
%%%%%%%%% i'm the ghost with the most, baby

\subsection{Background of Previous Observations}

As the nearest and one of the brightest RSGs, Betelgeuse ($\alpha$
Ori, M1--M2 Ia-Iab; Keenan \& McNeil 1989)\footnote{Not to be confused
  with Beetlejuice or the U.S.S.\ Betelgeuse.} is a prototypical
object for studying RSG mass loss.  Recent measurements suggest a
distance of 197$\pm$45 pc (Harper et al.\ 2008; see also Huggins 1987)
that we shall adopt here, although this is larger than (but consistent
with) the recently revised Hipparcos distance of 152$\pm$20 pc (van
Leeuwen 2007).  A detailed and up-to-date discussion of the distance
to Betelgeuse is given by Harper et al.\ (2008).

Because it is so nearby (and so cool and large), Betelgeuse has
afforded astronomers the opportunity to spatially resolve stellar and
atmospheric structure.  Indeed, Betelgeuse was the first star to be
spatially resolved interferometrically by Michelson \& Pease (1921),
with a diameter of about 44 mas.  More recent imaging and
interferometric techniques reveal that the apparent diameter of
Betelgeuse is about 57 mas (Bester et al.\ 1996), corresponding to a
radius of 5.6 AU or 1200 $R_{\odot}$.  Because of variability, an
average $T_{\rm eff}$ for Betelgeuse is $\sim$3300 K (Harper et al.\
2001)\footnote{This temperature of 3300~K is lower than the often
  quoted value of 3600~K, but is the most appropriate value for its
  $T_{eff}$ (see Harper et al.\ 2001; Ryde et al.\ 2006).}.  At a
distance of 152--197 pc, the star's luminosity is roughly
0.9--1.5$\times$10$^5$ L$_{\odot}$, implying an initial mass of 15--20
$M_{\odot}$.

The size measured in the mid-IR by Bester et al.\ (1996) is about 10\%
larger than optical measurements (Wilson et al.\ 1992), due to either
stronger limb-darkening at shorter wavelengths or effects of molecular
material above the photosphere (see Perrin et al.\ 2007).  Several
imaging and spectroscopic studies have suggested large-scale
non-uniform ``bright spots'' on the star's surface (Buscher et al.\
1990; Wilson et al.\ 1992; Josselin et al.\ 2007; Gray 2001, 2008),
reminiscent of suggestions made by Schwarzschild (1975) that large
RSGs may have only a few convective granulation cells with sizes
comparable to the stellar radius.  The chromospheric emission from
Betelgeuse has been imaged in the UV with {\it HST} (Gilliland \&
Dupree 1996; Uitenbroek, Dupree, \& Gilliland 1998), suggesting that
it is roughly two times larger in the UV than at visual wavelengths
with interesting departures from spherical symmetry.  High spatial
resolution spectroscopy with {\it HST} shows asymmetric and
time-variable infall and outflow across the star's disk (Lobel \&
Dupree 2001).  Radio and H$\alpha$ observations have also resolved
structure within a few stellar radii (Hebden et al.\ 1986; Skinner et
al.\ 1997; Lim et al.\ 1998).

The resolved structures most relevant to our 4.6~$\micron$ CO spectra
are more extended mid-IR features imaged at 10--20 $\micron$
associated with warm dust within $\sim$5\arcsec\ of the star (Hinz et
al.\ 1998; Rinehart et al.\ 1998).  These studies reveal mild
asymmetries in the dusty circumstellar environment.  The 10~$\micron$
nulling interferometer image from Hinz et al.\ (1998) is reproduced in
Figure~\ref{fig:imgOri}.  This extended dust traces the same spatial
region over which we detect extended 4.6~$\micron$ CO emission.

In addition to these small-scale structures that have been studied at
high spatial resolution, Betelgeuse also exhibits extended detectable
structure in its large-scale CSM.  This is due to its mass-loss rate
of $\dot{M}\simeq$2--4$\times$10$^{-6}$ $M_{\odot}$ yr$^{-1}$ (e.g.,
Harper \& Brown 2006; Harper et al.\ 2001; Glassgold \& Huggins 1986).
One of the most remarkable probes of this extended CSM is
resonant-scattered atomic emission, like K~{\sc i} $\lambda$7699
(Bernat \& Lambert 1976; Bernat et al.\ 1978; Honeycutt et al.\ 1980;
Mauron et al.\ 1984; Mauron 1990; Plez \& Lambert 2002), and Na~{\sc
  i}~D (Mauron \& Querci 1990; Mauron 1990; Mauron \& Guilain 1995).
The recent study of K~{\sc i} $\lambda$7699 by Plez \& Lambert (2002)
has been the most instructive, revealing multiple thin shells of
emission out to radii $\sim$55\arcsec\ from the star.  Although these
data reveal local inhomogeneities associated with clumps and thin
shells, they are consistent with basically spherical expansion of
10--15 km s$^{-1}$ out to large radii.  With $v_{exp}\simeq$15 km
s$^{-1}$, emission within $\sim$55\arcsec\ or 10$^4$ AU traces
material ejected by the star in the past few thousand years.
Betelgeuse's wind apparently collides with the surrounding ISM at much
larger projected radii $\sim$5\arcmin\ from the star (Noriega-Crespo
et al.\ 1997; Ueta et al. 2003).

%%%%%%%%%%%%%%%%%%%%%%%%% FIGURE 5 - Ori spectral plots %%%%%%%%%%%%%%%%%
\begin{figure}
\epsscale{0.98}
\plotone{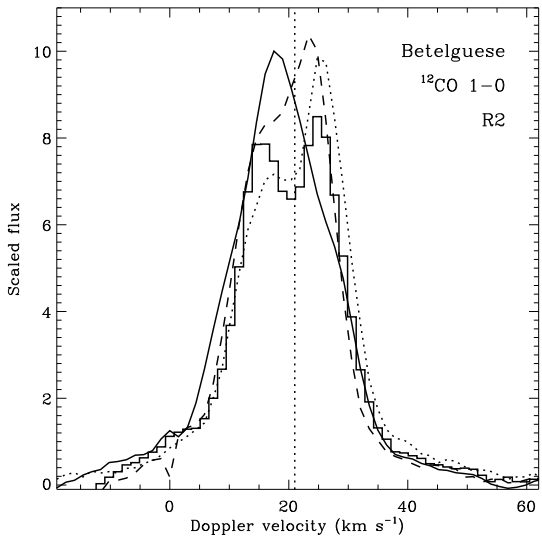}
\caption{Spectral tracings of the $^{12}$CO 1--0 R2 emission line
  profile in the shell of Betelgeuse at a few representative
  positions: N 0$\farcs$5 (solid), W 1\arcsec\ (dotted), W 1\arcsec\ N
  0$\farcs$5 (histogram), and W 1\arcsec\ N 1\arcsec\ (dashed).}
\label{fig:traceORI}
\end{figure}
%%%%%%%%%%%%%%%%%%%%%%%%%%%%%%%%%%%%%%%%%%%%%%%%%%%%%%%%%%%%%%%%%%%%%%

%%%%%%%%%%%%%%%%%%%%%%%%% FIGURE 6 - Ori radial plot %%%%%%%%%%%%%%%%%
\begin{figure}
\epsscale{0.98}
\plotone{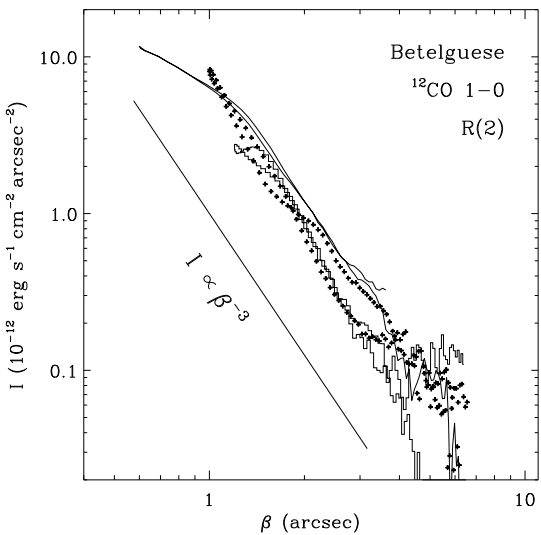}
\caption{Plots of the line surface brightness for $^{12}$CO 1--0 R2
  as a function of separation from the star $\beta$ for Betelgeuse.
  The CO emission was summed over velocities of $\pm$15 km s$^{-1}$
  compared to the systemic velocity, and intensity as a function of
  position along the slit was transformed to intensity as a function
  of separation from the star, taking into account the slit offset.
  Data for 3 example slit positions are shown: W 0$\farcs$5 (solid
  line), W 1$\farcs$0 (histogram), and S 1$\farcs$0 (plus signs).
  Also, intensity proportional to $\beta^{-3}$ is shown for
  comparison.}
\label{fig:radORI}
\end{figure}
%%%%%%%%%%%%%%%%%%%%%%%%%%%%%%%%%%%%%%%%%%%%%%%%%%%%%%%%%%%%%%%%%%%%%%

%%%%%%%%%%%
\subsection{Results from the CO Spectra}

Our 4.6~$\micron$ spectra of Betelgeuse reveal extended resonantly
scattered CO emission about 0$\farcs$5--3\arcsec\ from the star (see
the details of 1--0 $R$2 emission in Fig.~\ref{fig:velORI}).  This
corresponds to separations from the star of roughly 100--600 AU,
adopting a distance of 197 pc (Harper et al.\ 2008).\footnote{{\it
    Caveat}: When interpreting the position-velocity plots in
  Fig.~\ref{fig:velORI}, one must note that the stated offset slit
  position (commands given to the telescope) may not match the true
  offset position exactly due to potential pointing errors at the
  0$\farcs$1--0$\farcs$2 level, or real differences between the
  optical and IR centroids of the source.  This error is difficult to
  quantify, but it is presumably the explanation for why the offset
  positions at 0$\farcs$5 S and 0$\farcs$5 E are more severely
  contaminated by starlight than positions ostensibly located at the
  same offsets on the opposite sides of the star.  Alternatively,
  Fig.~\ref{fig:imgOri} shows that the star's centroid position is
  offset slightly to the S and E of the mid-IR centroid.}

From the same lines seen in absorption in the central star's spectrum,
Bernat et al.\ (1979) identified two shells, called S1 and S2, with
outflow speeds of 11 and 18 km s$^{-1}$, respectively.  Their true
spatial extent has not been determined since then.  We associate the
brightest 4.6~$\micron$ CO emission that we detect out to
$\sim$2\arcsec\ from the star with the slower S1 shell.  Fainter
emission from the faster S2 shell may be present at larger radii, but
the association of the CO emission with two separate and distinct
shells is not clear from our data.

Our Phoenix spectra reveal extended CO emission at velocities ranging
from $+$5 to $+$35 km s$^{-1}$ (Figs.~\ref{fig:velORI} and
\ref{fig:traceORI}), although most slit positions have emission
associated primarily with slower gas (the S1 shell).  The CO emission
centroid is at a heliocentric velocity of $+$21($\pm$1.5) km s$^{-1}$
(dashed line in Fig.~\ref{fig:velORI}), or $V_{\rm LSR}$=$+$5($\pm$1)
km s$^{-1}$.  These values are roughly consistent with those given by
Wilson (1953) measuring the star's photosphere, and with the centroids
of CO emission at $V_{\rm LSR}$=$+$3.7 km s$^{-1}$ (Kemper et al.\
2003; Huggins 1987) or C~{\sc i} emission at $+$3.0($\pm$1.3) km
s$^{-1}$ (Huggins et al.\ 1994).

From one slit position to the next, however, the peak and centroid can
shift to positive or negative velocities by $\pm$5 km s$^{-1}$, and
the profile shape can change.  Some positions (i.e.\
Figs.~\ref{fig:velORI}$a$, $b$, $g$, and $h$) reveal clear departures
from the intensity distribution one expects in a simple spherically
divergent steady wind.  Examples of extracted 1-D tracings of the line
profiles for a few positions are shown in Figure~\ref{fig:traceORI}.
This spatial variation in CO emission profiles cautions that
properties of the outflow deduced from absorption along our
sight-line, like the two discrete S1 and S2 shells, may not be
representative of the global wind.

First, at several positions the peak CO emission has a net blueshift
of $-$2 km s$^{-1}$ up to $-$5 km s$^{-1}$ compared to the systemic
velocity.  Second, at several positons we detect double- or
multiple-peaked emission profiles with central reversals
(Fig.~\ref{fig:velORI}h and \ref{fig:traceORI}).  Such profile shapes
evoke properties associated with thin shells, arcs, and discrete
clumps rather than a smooth and steady wind with $\rho\propto R^{-2}$.
The total variations in intensity associated with these features,
however, are typically quite small compared to the total emission ---
on the order of only $\sim$20\% of the total flux.

Structures seen when the P.A.=0\arcdeg\ slit was positioned 1\arcsec\
W of the star (Fig.~\ref{fig:velORI}$h$) showed the most successful
subtraction of stellar light, and reveal particularly intriguing
structure worthy of follow-up observations.  One can discern several
loops or arcs: one with a distinct central paucity of CO emission
centered roughly at the systemic velocity and with its spatial center
offset $\sim$0.5\arcsec\ N of the star (and 1\arcsec\ W), as well as
two more distant loop structures that also appear to have central
cavities but appear to have a net blueshift of roughly $-$5 km
s$^{-1}$.

From the spatial sampling of our data, however, it seems premature to
describe these structures as organized, large-scale departures from
spherical symmetry or discrete ejection events, although that surely
remains possible.  The data seem equally consistent with mild density
enhancements of factors as much as $\sim$20\% at certain positions,
signifying a large-scale but mild and random clumping of the wind on
size scales near our limiting spatial resolution of $\sim$50 AU.
Future studies with better spatial sampling, perhaps with an integral
field unit or Fabrey-Perot instrument, might be able to clarify the
structures seen here.

Overall, however, all CO emission features fall within a velocity
range of $\pm$15 km s$^{-1}$ of the assumed systemic velocity of $+$21
km s$^{-1}$ (heliocentric), consistent with the notion that the CO
structures correspond to mild density enhancements in a globally
spherical and nearly steady wind.  Ignoring local intensity peaks, the
overall intensity distribution falls roughly as $\beta^{-3}$, where
$\beta$ is the projected separation from the star (or impact
parameter).  This is demonstrated in Figure~\ref{fig:radORI}, where we
have taken intensity tracings along a few slit positions and
translated those positions along the slit to values of $\beta$ using
$\beta$=($x^2$+y$^2$)$^{0.5}$, where $x$ is the slit pointing offset,
and $y$ is the position along the slit for each pixel.

An intensity fall-off proportional to $\beta^{-3}$ is the value one
expects for a slice through an optically thin and steady wind with
$\rho \propto R^{-2}$. From 1\arcsec\ to about 3$\farcs$5 from the
star, the radial intensity distribution clearly follows $\beta^{-3}$.
Exterior to that point, the data are too noisy to provide useful
constraints, but there does appear to be a drop in intensity inside of
1\arcsec.  This may be due to higher optical depth, higher excitation
levels, molecule formation/survival rates, or other properties.

With the observed absolute fluxes we can calculate the mass-loss rate
from an analytic wind model, in which we assume a spherically
symmetric and homogeneous wind with a constant mass-loss rate, a
constant expansion velocity (both constant over a few hundred years),
and an optically thin wind in the CO lines along the
line-of-sight. The ratio of the wavelength-integrated, line-scattered
intensity, ${I_{CO,i}}$ (erg s$^{-1}$ cm$^{-2}$ arcsec$^{-2}$), and
the line-scattering flux $\bar {f}_{\lambda}$ (erg s$^{-1}$ cm$^{-2}$
cm$^{-1}$) as seen by the scattering molecules averaged across the
line width but measured at the distance $d$, is

\begin{eqnarray}
\label{equ1}
\frac {{\rm I_{CO,i}}(\beta)}{\bar {\rm f}_{\lambda}} & = & \frac{206,265}{32}\,    
\frac{e^2 \,\lambda ^2}{m_e \,c^2\,m_\mathrm H}\, f_{u\leftarrow l}\, \dot {M} 
 \times \, \nonumber \\
& & \frac{{N_i(\mathrm{CO})/N(\mathrm{CO})} \, \epsilon_{\mathrm{CO}}}{\mu\,v_\mathrm{exp} \, d}
\left (\frac{1}{\beta}\right )^3,
\end{eqnarray}
%%%%%%% this looks messy

\noindent where $f_{u\leftarrow l}=6.1\times 10^{-6}$ (Kirby-Docken \&
Liu 1978) is the absorption oscillator strength of the CO R2 line, and
$\epsilon_{\rm CO}=2.6\times 10^{-5}$ (Huggins et al.\ 1994) is the
fractional abundance of CO molecules relative to H,
i.e. $N$(CO)/$N$(H).  We assume this to be constant throughout the
envelope and that most oxygen is locked-up as CO molecules.  $N_i$(CO)
denotes the number density of CO molecules in the lower state,
i,\footnote{In the case of the CO R2 1--0 line this is the $^{12}$CO
  ($v''=0, J''=2$) level.} of the transition. As a mean, we assume
that 30\% of the CO molecules are excited to the $J$=2 level,
$N_i$(CO)/$N$(CO)$\simeq$0.3 (Ryde et al.\ 1999). Furthermore, $\mu$
is the mean molecular weight which we estimate to be $1.2$, $d$=197~pc
is the distance to Betelgeuse, $v_{\rm
  exp}=13.6\,\,\mbox{km\,s$^{-1}$}$ is the terminal expansion wind
velocity, and $\beta$ is the angular distance from the star on the sky
in seconds of arc.  The equation agrees well with the observed
$\beta^{-3}$ power dependence of the scattered intensity as a function
of the impact parameter on the sky, $\beta$ (see Fig.\ 6).  We see
directly that the mass-loss rate is linear with the scattered
intensity and expansion velocity by rewriting Eqn.~\ref{equ1} such
that

\begin{eqnarray} \label{mdot}
 \dot {M} & = & 4.20\times 10^{-10} \, \frac{v_\mathrm{exp} \, 
    d\,\beta^3}{N_i(\mathrm{CO})/N(\mathrm{CO}) \, \epsilon_{\mathrm{CO}}}  \times \nonumber \\
   & & \frac {{I_{CO,i}}(\beta)}{\bar{f}_{\lambda}} \, (M_\odot \, {\rm yr}^{-1}),
\end{eqnarray}

\noindent where $v_\mathrm{exp}$ is given in km s$^{-1}$, $d$ in pc,
$\beta$ in seconds of arc, ${I_{CO,i}}$ in
erg\,s$^{-1}$\,cm$^{-2}$\,arcsec$^{-2}$, and $\bar{f}_{\lambda}$ in
erg\,s$^{-1}$\,cm$^{-2}$\,$\mu$m$^{-1}$.  Thus, the mass-loss rate
deduced from the measured intensity at 2\arcsec\ away from the star
can be written as

\begin{equation}
\label{mdott}
\dot{M} = 1.15 \,\, {I_{\rm CO}}(\beta) / \bar{f}_{\lambda} \,\, (M_{\odot} \, yr^{-1}).
\end{equation}

\noindent From Figure~\ref{fig:radORI} we see that $I_{\rm
  CO}$($\beta$=2\arcsec) =
1.0($\pm$0.2)$\times$10$^{-12}$~erg\,s$^{-1}$\,cm$^{-2}$\,arcsec$^{-2}$,
while Justtanont et al.\ (1999) measure a total 4.6~$\mu$m flux from
Betelgeuse of $\bar{f}_{\lambda}\simeq$ 1.0$\times$10$^{-6}$
erg\,s$^{-1}$\,cm$^{-2}$\,$\mu$m$^{-1}$.  We then get roughly
1.2$\times$10$^{-6}$ $M_{\odot}$ yr$^{-1}$, with a likely uncertainty
in this absolute value of a factor of two in either direction, given
the number of different potential sources of error (real variations in
$I_{\rm CO}$ and calibration errors in $I_{\rm CO}$ due to slit
placement, uncertainty and time variability of $\bar{f}_{\lambda}$,
uncertainty in $N_i$(CO)/$N$(CO), etc.).  Despite these, this
independent estimate of the mass-loss rate is not far from the value
of 2--4$\times$10$^{-6}$ $M_{\odot}$ yr$^{-1}$ given by Harper et al.\
(2001); it is about half the lower bound, but the two values are
consistent given our likely uncertainties.  We would expect a much
larger discrepancy if our assumptions about the optically thin
emission or other factors were systematically wrong.

CO emission lines in a steady homogeneous wind of Betelgeuse should be
optically thin in the line-of-sight beyond

\begin{equation}
  p_{\tau=1} \simeq 1.9\times10^{28} \, \frac{N_i({\rm CO})}{N({\rm CO})}
  \frac{N({\rm CO})}{N({\rm H})} \, \frac{\dot{M}}{v_e^2} \, {\rm cm},
\end{equation}

\noindent where $\dot{M}$ is in units of $M_{\odot}$ yr$^{-1}$ and
$v_e$ is in km s$^{-1}$ (see Ryde et al.\ 1999, 2000).  With
$\dot{M}$=2$-$4$\times$10$^{-6}$, $\mbox{M$_\odot$\,yr$^{-1}$}$,
$N$(CO)/$N$(H)=2.6$\times$10$^{-5}$ (Huggins et al.\ 1994), and
$v_{\rm exp}=13.6\,\,\mbox{km\,s$^{-1}$}$ (Loup et al.\ 1993), the
impact parameter is $p_{\tau=1}$=1.6--3.2$\times 10^{15}$~cm, which
corresponds to 0$\farcs$45--1$\farcs$4 at a distance of 197$\pm$45~pc.
A radiative transfer model is needed for a more detailed
interpretation (Harper et al., in prep.).

Thus, aside from density fluctuations of $\pm$20\%, the CO spectra of
Betelgeuse are consistent with a basically spherical, optically thin
wind with a steady mass-loss rate over the past few hundred years.
VY~CMa presents a quite different picture, as we discuss next.

%%%%%%%%%%%%%%%%%%%%%%%%% FIGURE 7 - VY spec %%%%%%%%%%%%%%%%%%%%%%%%%
\begin{figure*}
\epsscale{0.95}
\plotone{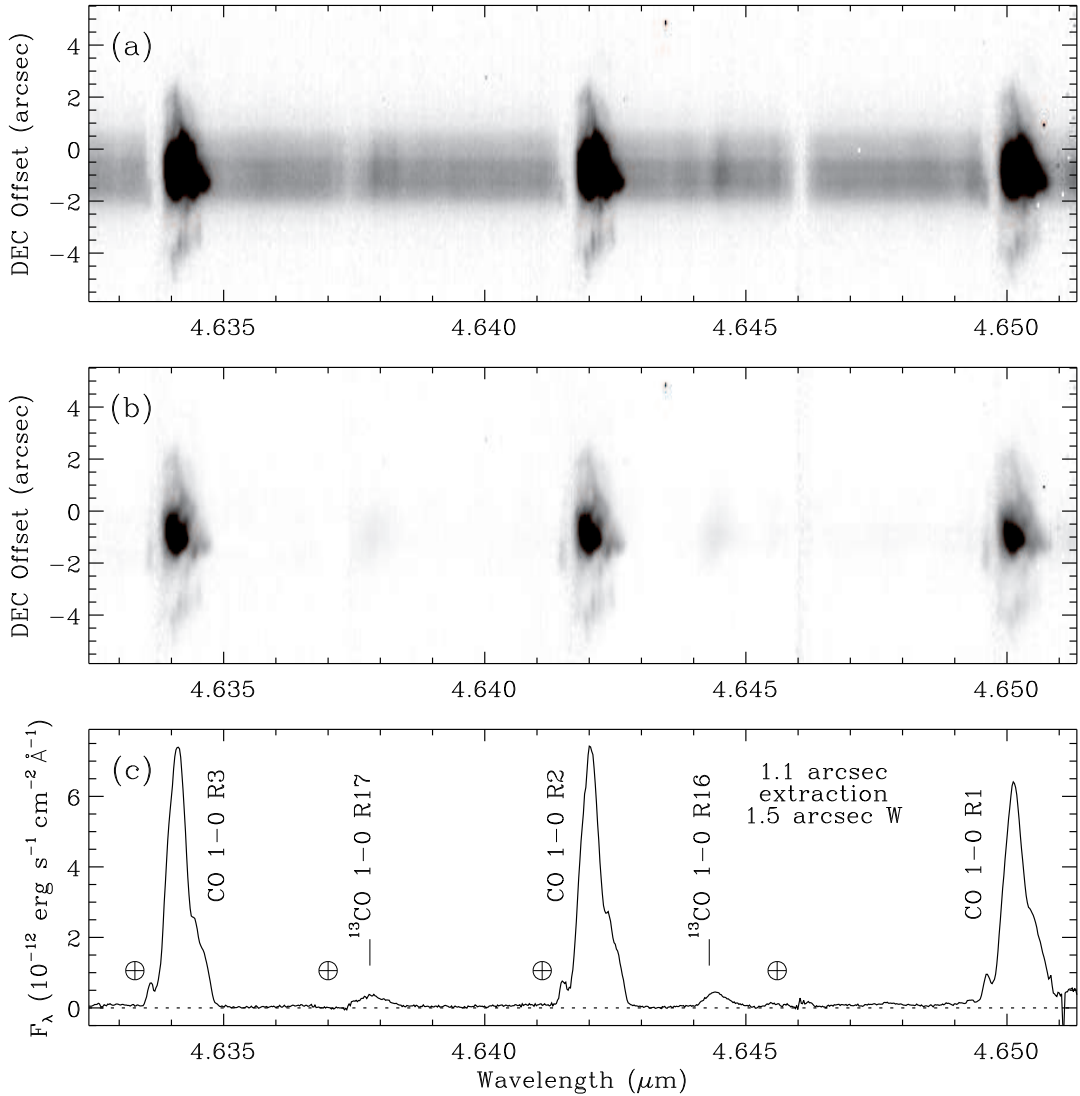}
\caption{Same as Figure~\ref{fig:fullORI} but for VY~CMa, with the
  slit at P.A.=0\arcdeg\ positioned 1$\farcs$5 west of the star.  This
  is an example of one of several slit positions.  In addition to the
  $^{12}$CO 1--0 lines, the 1--0 R17 and R16 transitions of $^{13}$CO
  are also clearly detected in these spectra, marked in the bottom
  panel.}
\label{fig:fullVY}
\end{figure*}
%%%%%%%%%%%%%%%%%%%%%%%%%%%%%%%%%%%%%%%%%%%%%%%%%%%%%%%%%%%%%%%%%%%%%%

%%%%%%%%%%%%%%%%%%%%%%%%% FIGURE 8 - VY stamps %%%%%%%%%%%%%%%%%%%%%%%
\begin{figure*}
\epsscale{0.95}
\plotone{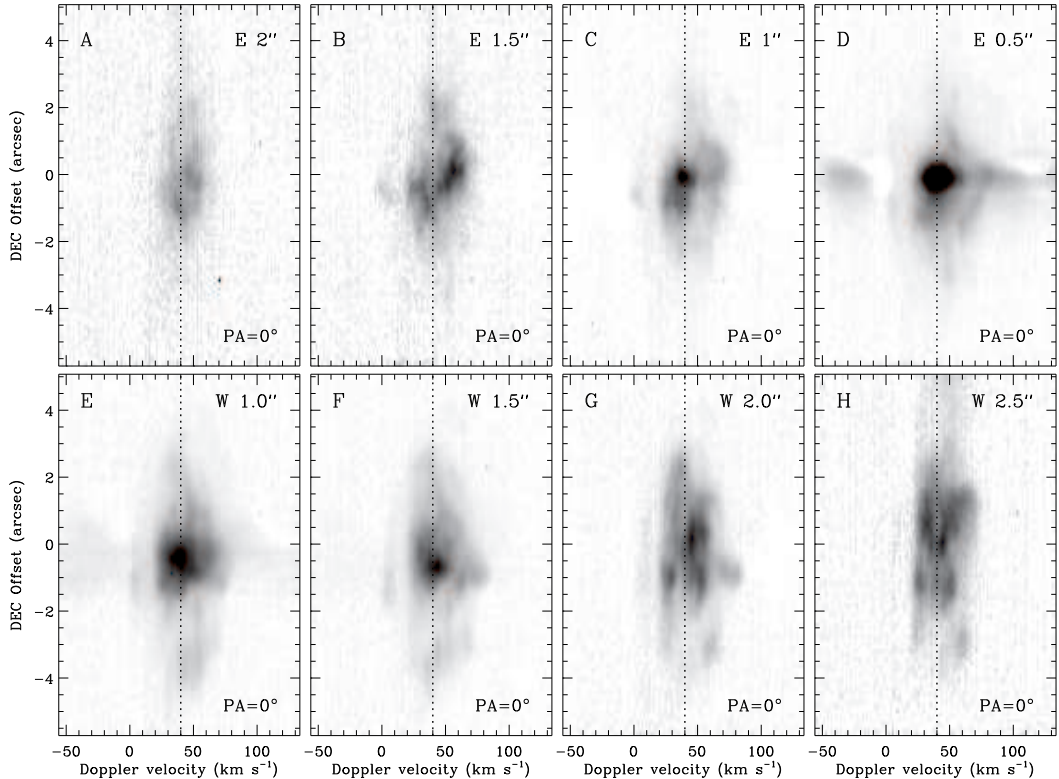}
\caption{Same as Figure~\ref{fig:velORI} but for VY CMa, for slits
  offset to the E and W of the star (see Fig.\ 2) with the slit along
  P.A.=0\arcdeg.  Slit offset positions from the star are indicated in
  each panel.  The systemic velocity of $+$40 km s$^{-1}$ (from Smith
  2004) is shown.}
\label{fig:velVYew}
\end{figure*}
%%%%%%%%%%%%%%%%%%%%%%%%%%%%%%%%%%%%%%%%%%%%%%%%%%%%%%%%%%%%%%%%%%%%%%

%%%%%%%%%%%%%%%%%%%%%%%%% FIGURE 9 - VY stamps %%%%%%%%%%%%%%%%%%%%%%%
\begin{figure*}
\epsscale{0.95}
\plotone{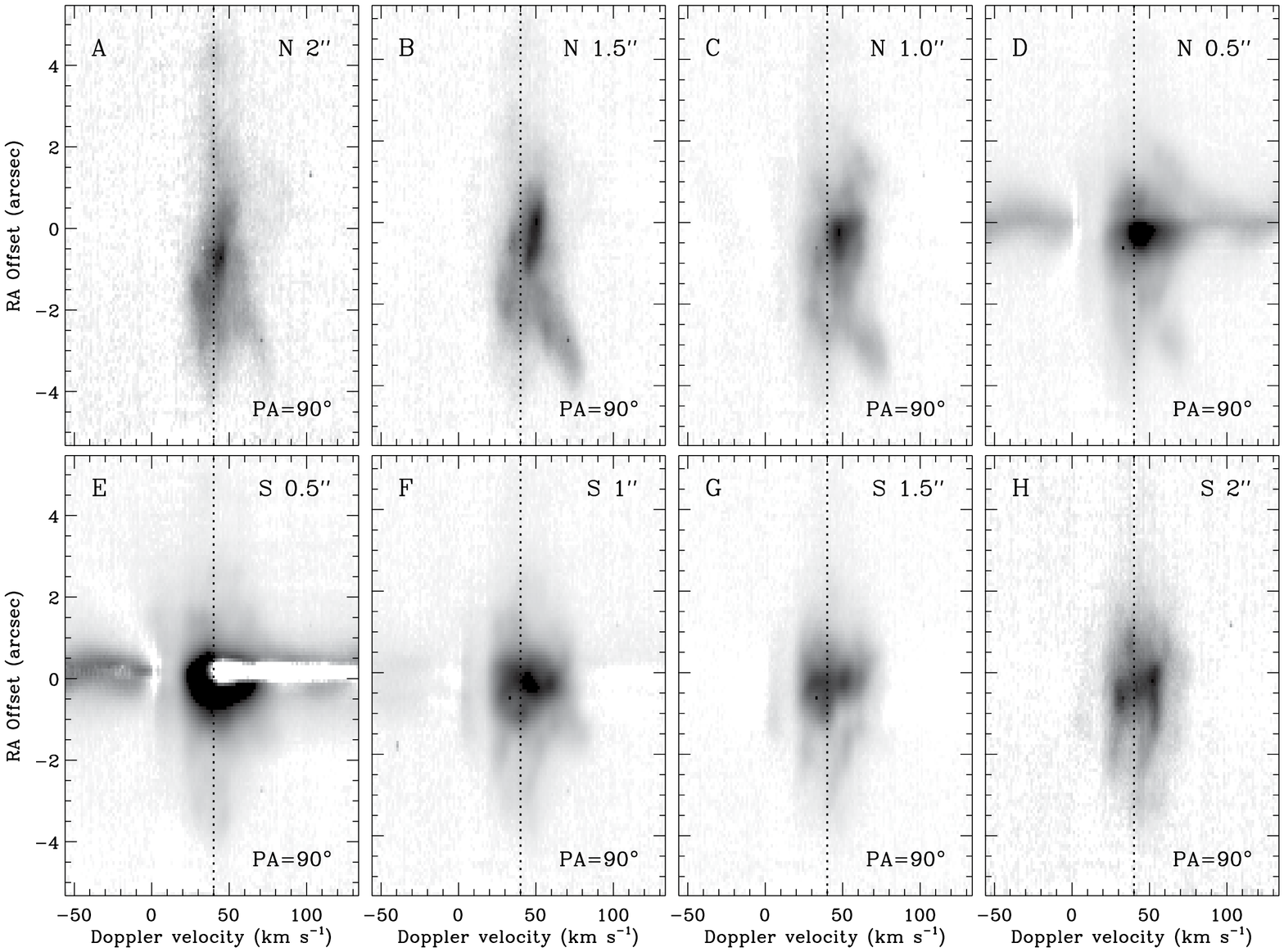}
\caption{Same as Figure~\ref{fig:velVYew}, for slits offset to the N
  and S of the star (see Fig.\ 2) with the slit along P.A.=90\arcdeg.
  Slit offset positions from the star are indicated in each panel. Up
  is toward the east.}
\label{fig:velVYns}
\end{figure*}
%%%%%%%%%%%%%%%%%%%%%%%%%%%%%%%%%%%%%%%%%%%%%%%%%%%%%%%%%%%%%%%%%%%%%%

\section{RESULTS AND DISCUSSION FOR VY~CMa}

\subsection{Background of Previous Observations}

Like Betelgeuse, VY~Canis~Majoris (M2.5--M5e~Ia; e.g.\ Wallerstein 1958)
is bright and has been a favorite target of astronomers interested in
RSG mass loss.  We adopt a distance of 1.5~kpc (Herbig 1972; Marvel
1997), although there is still some disagreement in the literature
regarding the distance to VY~CMa.  For example, Choi et al.\ (2008)
give an estimate of 1.14 kpc.

VY~CMa is distinguished from Betelgeuse in that it is intrinsically
more luminous and has a much higher mass-loss rate.  At 1.5 kpc, the
integrated IR luminosity from 0.4 to 25~$\micron$ is
4.1$\times$10$^{5}$ L$_{\odot}$ or $M_{Bol}$=$-$9.3 mag (Smith et al.\
2001), which is a minimum luminosity for the system because some
optical luminosity may escape in directions away from our
line-of-sight and may not be absorbed and re-radiated by dust.  This
suggests an initial mass of roughly 35 $M_{\odot}$ (Smith et al.\
2001).  (If the distance is as low as 1.14 kpc [Choi et al.\ 2008],
then the implied initial mass would be $\sim$25 $M_{\odot}$.)
VY~CMa's current mass-loss rate has been estimated as
2--4$\times$10$^{-4}$ $M_{\odot}$ yr$^{-1}$ (Danchi et al.\ 1994),
about 10$^2$ times higher than Betelgeuse, although the star may have
had an even higher mass-loss rate of $\sim$10$^{-3}$ $M_{\odot}$
yr$^{-1}$ during episodes in its recent history (Smith et al.\ 2001).
VY~CMa also has a faster wind speed of 30--40 km s$^{-1}$ (e.g., Reid
\& Dickenson 1976; Reid \& Muhleman 1978; Zuckerman \& Dyck 1986),
compared to $\sim$15 km s$^{-1}$ for Betelgeuse.  The slower speeds of
Betelgeuse are more typical of RSGs.

Based on atmospheric models for the observed visual spectrum at
spectral type M2.5~Ia, Massey et al.\ (2006) have recently suggested
that VY~CMa's effective temperature is around 3650~K, higher than
previous claims.\footnote{Note, however, that the lower luminosity of
  6$\times$10$^4$ L$_{\odot}$ suggested by Massey et al.\ conflicts
  with the observed total IR luminosity (Smith et al.\ 2001).  Also,
  see comments by Humphreys (2006) regarding the possible effects of
  spectral-type variation (M2.5 vs.\ M4--5) and heavy mass loss on the
  $T_{eff}$ determination.} With the luminosity above, this implies a
stellar radius roughly 30--40\% larger than that of Betelgeuse.
Massey et al.\ (2006) pointed out that cooler temperature estimates in
the literature would violate hydrostatic stability posed by the
Hayashi limit.  Yet, the observed variations from M2.5 to later
spectral types and cooler temperatures are observed, so one could
hypothesize that violating the Hayashi limit of hydrostatic and
convective stability --- at least for limited periods in VY~CMa's
variability cycle --- might {\it cause} periods of enhanced episodic
mass loss, spewing material toward random directions (Smith et al.\
2001; Smith 2004; Humphreys et al.\ 2005, 2007).  Thus, temporary
violations of the Hayashi limit may be a clue to VY~CMa's instability
and mass loss.  We will return to this later.

VY~CMa is unusual among RSGs in that it displays a prominent
asymmetric red reflection nebula $\sim$10\arcsec\ in size, which has
been known for almost a century (Perrine 1923) and can be seen even in
small telescopes.  Visual-wavelength imaging with {\it HST} revealed
complex structure in the nebula, with many clumps and arc structures
that do not adhere to any spherical symmetry or axisymmetry (Smith et
al.\ 2001).  {\it HST} imaging also revealed that the central star is
highly obscured because its position moves with wavelength:
blue/near-UV images see primarily reflected light, while red and
near-IR wavelengths begin to penetrate the dust as the central point
source shifts to the northeast (Smith et al.\ 2001; Kastner \&
Weintraub 1998).  Color maps of the reflection nebula reveal a
large-scale extinction gradient such that sight-lines through the
nebula northeast of the star tend to suffer considerably more
reddening and extinction (Smith et al.\ 2001).  This circumstellar
dust absorbs most of the star's photospheric luminosity, reradiating
that energy at near- to mid-IR wavelengths.  VY CMa's SED has been
studied extensively at IR wavelengths, revealing a strong but
apparently temporally variable IR excess, plus a variable
9.7~$\micron$ silicate feature (Gehrz et al.\ 1970; Gillett et al.\
1970; Herbig 1970; Merrill \& Stein 1976a, 1976b; Danchi et al.\ 1994;
Le Sidaner \& Le Bertre 1996; Monnier et al.\ 1998, 2000; Smith et
al.\ 2001).

VY CMa is also an extended object in the IR.  Near-IR images of the
nebula presented by Monnier et al.\ (1999) showed remarkable structure
very close to the obscured star (see also Monnier et al.\ 2004), some
of which is hidden in visual {\it HST} images.  Earlier near-IR images
suggested that the central object was resolved and non-spherical on
sub-arcsecond scales (Wittkowski et al.\ 1998; Bensammer et al.\
1985), while fainter extended near-IR emission has also been detected
out to $\sim$5\arcsec\ from the star (Monnier et al.\ 1999; Smith et
al.\ 2001).  The central object appears elongated along an east/west
axis at mid-IR wavelengths (Lipscy et al.\ 2005; Smith et al.\ 2001),
which is different from the extension at shorter near-IR wavelengths
where a ``bubble'' appears to extend 0$\farcs$1--0$\farcs$2 south from
the central source (Monnier et al.\ 2004).  All this suggests
time-variable and direction-dependent ejection of material by the
central star.  Yet, there are some hints of a persistent N-S or NE-SW
axis of symmetry (Monnier et al.\ 1999, 2004; Smith et al.\ 2001).

VY~CMa is also a famous resolved OH, H$_2$0, and SiO maser source
(Wilson \& Barrett 1968; Eliasson \& Bartlett 1969; Snyder \& Buhl
1975; Van Blerkom \& Auer 1976; Rosen et al.\ 1978; Benson \& Mutel
1979, 1982; Deguchi et al.\ 1983; Bowers et al.\ 1983, 1993; Marvel et
al.\ 1997; Zheng et al.\ 1998; Richards et al.\ 1998; Shinnaga et al.\
2004).  The results of these numerous studies are too complex and
varied to review in detail here, but they paint a picture of a
structured molecular outflow within $\sim$1000 AU of the star
expanding at 30--40 km s$^{-1}$.  Maser motions also provide
indications of possible axial symmetry, as we discuss below.

\subsection{K~{\sc i} and CO Emission}

Extended K~{\sc i} $\lambda$7699 emission has been seen in a number of
other RSGs besides Betelgeuse, such as $\alpha$~Her and Mira (Mauron
\& Caux 1992; Gustafsson et al.\ 1997).  However, VY CMa is the only
Galactic RSG in which bright K~{\sc i} emission lines appear in the
spectrum of the star itself.  This was noted long ago by Wallerstein
(1958), and the unusually strong K~{\sc i} emission lines --- plus
numerous other extremely unusual {\it emission} lines like TiO, ScO,
Ti~{\sc i}, Cr~{\sc i}, Rb~{\sc i}, Ba~{\sc ii}, etc.\ --- have been
studied previously (Joy 1942; Wallerstein 1958, 1986; Hyland et al.\
1969; Wallerstein \& Gonzalez 2001).

The spatially extended nature of the K~{\sc i} emission around VY~CMa
was first noted by Smith (2004), who found that the resonant-scattered
emission followed the spatial distribution of the visual reflection
nebula, and could be used to trace its kinematics.  Smith (2004) found
that the kinematics gave strong evidence for asymmetric and episodic
mass loss due to localized eruptions on the stellar surface through
discrete ejection events in the past $\sim$1000 yr. Similar long-slit
spectra of K~{\sc i} emission were presented subsequently by Humphreys
et al.\ (2005), who confirmed these results and also emphasized the
evidence for discrete ejection events from the kinematics of
individual nebular features.  More recently, these K~{\sc i} Doppler
kinematics were combined with {\it HST} measurements of proper motions
and polarization to disentangle the 3-D orientation of specific
features and to estimate their true (deprojected) velocities and ages
(Humphreys et al.\ 2007; Jones et al.\ 2007).  These results
underscored the importance of individual ejection events over the past
100--1000 yr, with no clear indication of axial symmetry.

This evidence for asymmetry seems to contradict several studies
mentioned earlier. Molecular and maser studies at longer wavelengths
generally point toward some degree of bipolar or toroidal symmetry in
the nebula (Muller et al.\ 2007; Shinnaga et al.\ 2004; Richards et
al.\ 1998; Bowers et al.\ 1983; Deguchi et al.\ 1983; Benson \& Mutel
1982, 1979; Rosen et al.\ 1978), as well as the IR SED interpreted as
a thick disk by Herbig (1970), the resolved IR structures near the
star (Monnier et al.\ 1999, 2004), and the a large-scale extinction
gradient across the nebula increasing from SW to NE (Smith et al.\
2001).

The kinematics of 4.6~$\micron$ CO emission we present below
underscore the prevalence of episodic mass ejections in random
directions inferred from K~{\sc i} Doppler shifts, visual continuum
proper motions, and visual polarization.  However, our CO kinematics
also add new information, because at 4.6~$\micron$, the CO emitting
clumps suffer less line-of-sight extinction and are therefore less
biased toward detecting features on the near side of the nebula.  This
turns out to be important, as we show next.

%%%%%%%%%%%%%%%%%% FIGURE 10 - VY spectral tracings %%%%%%%%%%%%%%%%%
\begin{figure}
\epsscale{0.98}
\plotone{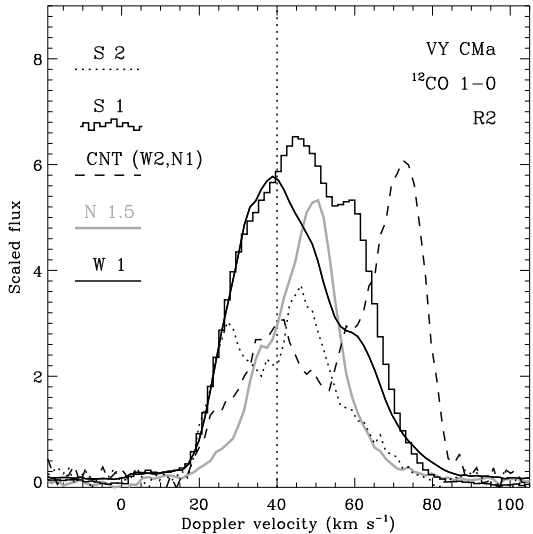}
\caption{Same as Fig.~\ref{fig:traceORI}, but for VY~CMa.  Example
  tracings of the CO line profile for: 2\arcsec\ south (amid the
  bubble outlined by ``arc 2'' in Fig.~\ref{fig:img}), a bright clump
  1\arcsec\ south (``S'' in Fig.~\ref{fig:img}), a position in the
  ``curved nebulous tail'' (2\arcsec\ west and 1\arcsec\ north), a
  position 1$\farcs$5 north of the star that is inconspicuous in
  visual images, and in the bright ejecta 1\arcsec\ west of the star.
  VY~CMa's systemic velocity of $+$40 km s$^{-1}$ is shown with a
  dotted line.  The blue wings of all profiles are truncated by
  self-absorption.}
\label{fig:traceVY}
\end{figure}
%%%%%%%%%%%%%%%%%%%%%%%%%%%%%%%%%%%%%%%%%%%%%%%%%%%%%%%%%%%%%%%%%%%%%%

%%%%%%%%%%%%%%%%%%%%%%%%% FIGURE 11 - VY radial plot %%%%%%%%%%%%%%%%%
\begin{figure}
\epsscale{0.98}
\plotone{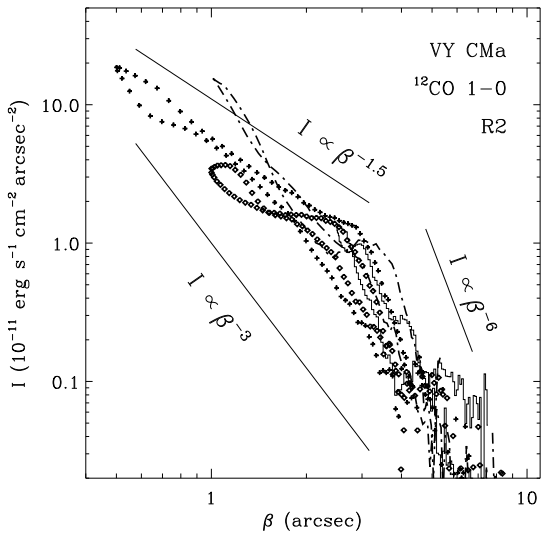}
\caption{Same as Fig.~\ref{fig:radORI}, but for VY~CMa. The CO
  emission was summed over velocities of $\pm$40 km s$^{-1}$ from the
  systemic velocity.  Data for four example slit positions are shown:
  W 1$\farcs$0 (dot-dash), W 2$\farcs$5 (histogram), N 0$\farcs$5
  (plus signs), and N 1$\farcs$0 (diamonds).  For comparison,
  intensity profiles proportional to $\beta^{-3}$, $\beta^{-1.5}$, and
  $\beta^{-6}$ are shown.}
\label{fig:radVY}
\end{figure}
%%%%%%%%%%%%%%%%%%%%%%%%%%%%%%%%%%%%%%%%%%%%%%%%%%%%%%%%%%%%%%%%%%%%%%

\subsection{Results from the CO Spectra}

Our new 4.6~$\micron$ spectra of the nebula of VY~CMa reveal the
kinematics and spatial distribution of CO gas with the highest
combination of spatial and spectral resolution yet, and they are the
first detection of extended emission in the IR vibration-rotation
lines of CO.  Figures~\ref{fig:velVYew} and \ref{fig:velVYns} show
detailed position-velocity plots of CO 1--0 $R$2 emission at radii
ranging from 0$\farcs$5--5\arcsec\ from the star.  In terms of spatial
and spectral resolution, our study improves significantly upon recent
efforts to discern the spatio-kinematic structure of VY~CMa's
molecular envelope, especially in CO, using mm wavelength observations
(e.g., Muller et al.\ 2007).

Examining the long-slit spectra in Figures~\ref{fig:velVYew} and
\ref{fig:velVYns} leads to several qualitative comments regarding the
observed structure (Fig~\ref{fig:img}a gives names of specific
features):

(1) The most important result is that in these 4.6~$\micron$ data, the
nebula appears more symmetric than it does in optical and near-IR
images (Monnier et al.\ 1999; Smith et al.\ 2001) or K~{\sc i} spectra
(Smith 2004; Humphreys et al.\ 2005).  Individual CO knots are
distributed asymmetrically with no clear signs of axial or point
symmetry (i.e. any individual condensation is not mirrored by a
companion on the opposite side of the star), but the global
distribution of CO condensations and diffuse emission is more
isotropic, following the general trend of increasing isotropy with
increasing wavelength (Smith et al.\ 2001).  The CO emission is still
brighter toward the S and W of the star, but the asymmetry is not
nearly as severe as at visual wavelengths.  We detect more redshifted
emission than previous K~{\sc i} studies, as well as relatively
brighter emission to the N and E of the star.

(2) All slit positions show pervasive blueshifted P Cygni absorption
at +15 km s$^{-1}$ (heliocentric), indicating a relatively smooth and
isotropic outflow at $-$25 km s$^{-1}$ with respect to the star.  This
causes all line profile tracings to have a blunt edge on the
blueshifted side.  This absorption is intrinsic to the source --- it
is different from the atmospheric CO absorption at 0 to --10 km
s$^{-1}$ (heliocentric).  This outflow speed of 25 km s$^{-1}$ is
slower than the usually adopted value of 30--40 km s$^{-1}$ indicated
by masers and the motions of dense condensations in CO (this work) and
K~{\sc i} (Smith 2004).

(3) There is some asymmetry in the {\it character} of the CO emission
distribution.  Namely, CO emission tends to be concentrated in more
isolated clumps to the S and SW of the star, and the structure seems
smoother toward the E and N.  This is especially apparent as one moves
progressively from E to W in panels $a$--$h$ in
Figure~\ref{fig:velVYew}.

(4) The feature in images known as the curved nebulous tail (CNT) has
CO emission that is predominantly redshifted, and shows Hubble-like
flow with steadily increasing Doppler shift with separation from the
star.  Maximum Doppler shifts of +40 km s$^{-1}$ with respect to the
star (+80 km s$^{-1}$ heliocentric) are attained at offsets 4\arcsec\
W and a few arcseconds N.  Like the Hubble flow seen in K~{\sc i} from
arc 1 (Smith 2004), this is a clear indication that the structure
formed as the result of an asymmetric and episodic mass ejection from
the star.  In CO spectra, the CNT is the most distinct feature in the
nebula.  Spectral tracings of line profiles for various nebular
features in Figure~\ref{fig:traceVY} show that the CNT stands out as
unique compared to other clumps in the nebula.

(5) At positions several arcseconds S and 1--2\arcsec\ W of the star,
with N/S-oriented slits running through the large bubble known as
``arc 1'', little blueshifted CO emission is seen.  This is in stark
contrast to the kinematics seen in K~{\sc i}, where one sees a
blueshifted Hubble-like flow increasing in speed to --30 km s$^{-1}$
with respect to the systemic velocity (Smith 2004).  This is one of
the most prominent features in visual images of the nebula, and it
appears to be absent in CO data.  Our spectra missed the limb of arc
1, so perhaps the CO emission is concentrated at its edge, but the CO
distribution must be different from that of K~{\sc i}, which fills the
bubble's interior (Smith 2004).

(6) In general, we find that reddened features (those with a high
F1042M to F410M flux ratio in {\it HST} images; Fig.\ 9b in Smith et
al.\ 2001) have redshifted Doppler velocities.  This includes diffuse
emission NW of the star, arc 2, clump SW, and the CNT.

(7) There is surprising redshifted CO structure located 1--2\arcsec\ E
and N of the star (examine Figs.~\ref{fig:velVYew}e and
\ref{fig:velVYew}f with the slits oriented N/S, as well as
Fig.~\ref{fig:velVYns}d with the slit oriented E/W), which is not seen
in K~{\sc i} data (Smith 2004): Namely, on the back side of the
nebula, dense clumps of CO-emitting material appear to be piling-up in
a redshifted ``ridge'' of emission.  This corresponds to the region of
highest average reddening in the nebula, and may be related to a
feature designated as arc 3 in visual {\it HST} images (see Smith et
al.\ 2001).  This feature is difficult to see even with {\it HST}
because it is faint and reddened, and nearly overwhelmed by the glare
of the central star. It is interesting to speculate that this putative
``pile-up'' on the E and back side of the nebula may be related to a
density gradient in VY~CMa's interstellar environment, since the
molecular cloud L1667 is located a few arcminutes to the east of
VY~CMa.

Altogether, spectral intensity tracings of CO emission around VY~CMa
(Fig.~\ref{fig:traceVY}) reveal an environment that is quantitatively
different from that of Betelgeuse (Fig.~\ref{fig:traceORI}).  CO
emission profiles around Betelgeuse show centrally peaked and nearly
gaussian profiles at all slices through the shell, with only minor
fluctuations due to clumping at the 10--20\% level.

VY~CMa, in contrast, has huge variation from one position to the next
(Fig.~\ref{fig:traceVY}), including: (1) wide but centrally
concentrated profiles in the dense environment close to the star
(1\arcsec\ W and 1\arcsec\ S), (2) narrower and irregular
multiple-peaked CO profiles (1$\farcs$5 N and 2\arcsec\ S), and (3)
strong emission peaks offset far from the systemic velocity by up to
40 km s$^{-1}$, as in the CNT.  These are not the profiles one expects
for slices through a steady wind, suggesting that the vast majority of
4.6~$\micron$ CO emission arises in the densest cloudlets rather than
from material associated with an underlying homogeneous wind.

Similarly, spatial intensity tracings of VY~CMa's CO emission are
inconsistent with a steady optically thin wind.  As noted above in the
discussion of Betelgeuse, one expects the apparent intensity to vary
as $\beta^{-3}$ (where $\beta$ is the projected angular separation
from the star) for slices through a steady wind with density falling
as $R^{-2}$.  Betelgeuse follows this trend (Fig.~\ref{fig:radORI}),
but VY~CMa does not.  Spatial intensity tracings of VY~CMa are shown
in Figure~\ref{fig:radVY}.  Instead of $I\propto\beta^{-3}$, VY~CMa
shows a much shallower intensity drop out to $\sim$3\arcsec\ from the
star, and then a much {\it steeper} profile closer to
$I\propto\beta^{-6}$ from 3\arcsec--5\arcsec\ separations.  Beyond
5\arcsec\ the profile may return to $I\propto\beta^{-3}$, but the data
become noisy.

The observed intensity variations with projected radius indicate
strong variations in the mass-loss rate during VY~CMa's recent
history.  This assumes optically thin $^{12}$CO emission, which we
demonstrate in the following section by a comparison between $^{12}$CO
and $^{13}$CO line profiles. The observed deviation of CO intensity
from the normal $\beta^{-3}$ profile in Figure~\ref{fig:radVY} could,
in principle, be caused by systematic changes in outflow velocity with
a steady mass-loss rate.  We find this unlikely, however. The enhanced
emission at $\beta\simeq$3\arcsec\ from the star, indicating higher
densities than in a steady wind at the same projected radius, would
require outflow speeds that are a factor of 5--10 {\it slower} than
the normal wind, representing a systematic deceleration of VY~CMa's
wind to well below its escape velocity.  This possibility is ruled out
by the observed kinematics in our CO spectra, which show no trend of
slow gas concentrated at these large radii.  In fact, we see some
indications of the opposite trend, with higher speeds at large radii.
This therefore requires a surge of mass-loss in VY CMa's recent past.

Assumming outflow at 35 km s$^{-1}$, the steep $I\propto\beta^{-6}$
profile from 3--5\arcsec\ implicates a sharp rise in the mass-loss
rate $\sim$1000~yr ago that peaked about 600~yr ago, and gradually
returned to its present state thereafter.  To explain the observed
change in CO emission intensity, the increase would have needed to be
a factor of 5--10 above its current average mass-loss rate of
2--4$\times$10$^{-4}$~$M_{\odot}$ yr$^{-1}$ (Danchi et al.\ 1994).
This implies an average mass-loss rate 600-1000 years ago of roughly
2$\times$10$^{-3}$~$M_{\odot}$ yr$^{-1}$.  This is in line with a
previous estimate from bright dust condensations seen in the
reflection nebula (Smith et al.\ 2001), indicating that the CO and
reflecting dust trace the same dense condensations, and that these
cloudlets contain most of the emitting mass in the nebula.  Our
conclusion of variable mass loss in the past is also in general
agreement with recent results from theoretical modeling of unresolved
mm-wavelength CO line profiles by Decin et al.\ (2006), who required a
strong increase in the mass-loss rate about 1000 yr ago, with lower
$\dot{M}$ before and after, in order to explain the multiple-peaked
line profile shapes in VY~CMa.

Our 4.6~$\micron$ CO spectra, however, spatially resolve individual
features resulting from the enhanced mass loss, showing that it was
not simply an increase in the spherical time-averaged wind $\dot{M}$
of the star, but rather, that the enhanced mass loss consisted of
numerous discrete ejections from the star's surface.  While
4.6~$\micron$ CO spectra reveal that the {\it distribution} of ejecta
is nearly isotropic --- or at least far less asymmetric than implied
by visual-wavelength images suffering from extinction --- we stress
that the ejection of any {\it individual} condensation corresponding
to an episodic event is highly asymmetric.  Thus, the physics behind
this type of mass ejection and the increased average mass-loss rate of
VY~CMa 600--1000 yrs ago is likely connected to pulsational,
convective, or magnetohydrodynamical activity in the outer layers of
the star, rather than the physics of steady wind driving in the outer
atmosphere.  In this context, where VY~CMa apparently suffers
temporary episodes of increased mass loss and increased instability,
its proximity to (and its possible violation of) the Hayashi limit for
convective stability is perhaps not so surprising.  In fact, increased
instability and mass loss is the likely way a star would respond to
being pushed past the Hayashi limit in order to re-establish
equilibrium.  The possible link between discrete mass ejection events
and individual giant convection cells is intriguing.

%%%%%%%%%%%%%%%%%%%%%%%%% FIGURE 12 - 12co - 13co %%%%%%%%%%%%
\begin{figure}
\epsscale{0.93}
\plotone{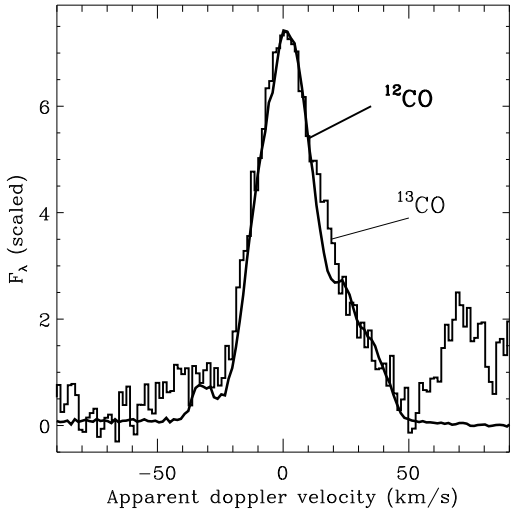}
\caption{Line profile of $^{12}$CO 1--0 R2 $\lambda$46412 (thick solid
  line) compared to that of the adjacent $^{13}$CO 1--0 R16 line (thin
  histogram).  The intensity of the $^{13}$CO line is multiplied by a
  factor of 16.4 to match that of the $^{12}$CO line.}
\label{fig:13co}
\end{figure}
%%%%%%%%%%%%%%%%%%%%%%%%%%%%%%%%%%%%%%%%%%%%%%%%%%%%%%%%%%%%%%%%%%%%%%

\subsection{$^{13}$CO in VY CMa}

Our spectroscopy of VY~CMa's nebula also provides clear detections of
faint lines of $^{13}$CO emission (see Fig.~\ref{fig:fullVY}c).
Comparing $^{12}$CO 1--0 $R$2 to the $^{13}$CO 1--0 R16 line yields a
measured $^{12}$CO/$^{13}$CO intensity ratio of roughly 16.4.
Converting this observed intensity ratio to an intrinsic
$^{12}$C/$^{13}$C isotopic ratio requires a detailed model of the gas
excitation temperature and observations of additional lines.  However,
we can place rough constraints on VY~CMa's isotopic ratio by
comparison to Betelgeuse, which has a well-established
$^{12}$C/$^{13}$C isotopic ratio of $\sim$7, and for which we observe
an intensity ratio of no less than 19.0 for $^{12}$CO 1--0 $R$2 to
$^{13}$CO 1--0 R16 from the spectrum in Fig.~\ref{fig:fullORI}.
Assuming that the CO excitation temperatures in VY CMa and Betelgeuse
are not wildly discrepant, this then implies that VY~CMa has an
intrinsic $^{12}$C/$^{13}$C isotopic ratio of $\la$6.  This is far
below the solar-system ratio of $\sim$90, showing that significant
quantities of nuclear-processed material containing $^{13}$C have been
dredged-up into VY CMa's envelope, as expected in a post-main-sequence
star.

Figure~\ref{fig:13co} compares profiles of the $^{12}$CO and $^{13}$CO
lines from the spectrum in Fig.~\ref{fig:fullVY}c. The profiles of
$^{12}$CO and $^{13}$CO match almost perfectly at all velocities.  The
only significant difference between the two lines is at velocities
near --40 km s$^{-1}$ (relative to the peak), where the $^{12}$CO line
is more severely affected by telluric absorption.  Given the very
different strengths of the two lines, this similarity confirms that
the $^{12}$CO emission is indeed optically thin, even for the brighter
and slower gas closest to the star.  This is important because it
justifies our previous assumption (\S 4.3) of optically thin $^{12}$CO
emission when diagnosing the mass-loss history of VY~CMa.

%%%%%%%%%%%%%%%%%%%%%%%%% FIGURE 13 - VY cartoon %%%%%%%%%%%%%%%%%
\begin{figure}
\epsscale{0.99}
\plotone{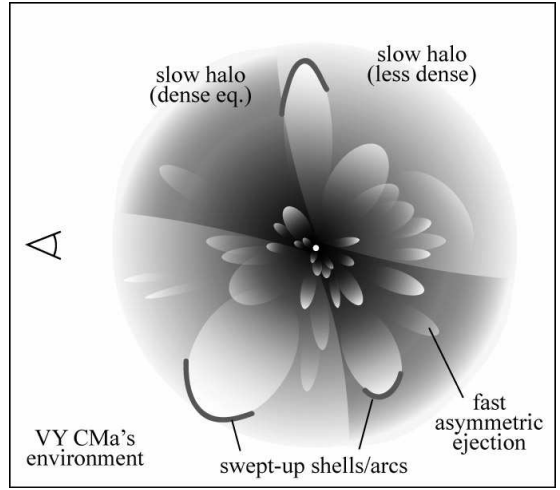}
\caption{Cartoon of the proposed schematic geometry and structure in
  the likely pre-SN environment around VY~CMa consisting of individual
  asymmetric mass ejections embedded within a larger and axisymmetric
  slow envelope (see text). An Earth-based observer is to the left,
  such that a region of relatively low density is tilted toward us and
  is seen primarily south of the star.  Maser emission probably arises
  where the innermost ejections sweep into the dense ``equatorial''
  zones.}
\label{fig:cartoon}
\end{figure}
%%%%%%%%%%%%%%%%%%%%%%%%%%%%%%%%%%%%%%%%%%%%%%%%%%%%%%%%%%%%%%%%%%%%%%

\subsection{A Geometric Model for VY~CMa's Nebula}

How can we reconcile the obviously discrepant observational results
where (1) visual/near-IR images and spectra of the densest features
show a very asymmetric nebula with no clear signs of spherical or
axial symmetry, and (2) maser and other molecular studies generally
suggest axial symmetry with a preferred axis oriented NE/SW, with
systematic extinction gradients along the same axis?  The new CO
spectra we present here may provide a useful bridge between these two,
because while there is an overall sense of nearly isotropic outflow in
terms of the speed and distance from the star, there is also clear
evidence that each {\it individual} clump embedded in that outflow was
the result of a highly asymmetric mass-loss event.

We propose the following basic picture for the outflow around VY~CMa:
A lower density and relatively slow ``halo'' (e.g., Smith et al.\
2001) is punctuated with faster and denser CO/dust cloudlets that
result from episodic mass ejections, as depicted schematically in
Fig.~\ref{fig:cartoon}.  Bernat (1981) found similar evidence for
multiple ejections from CO absorption profiles in a number of RSGs.

The halo is probably responsible for the ubiquitous 25 km s$^{-1}$
outflow indicated by blueshifted CO absorption features.  If it were
to contain a mild sense of axial symmetry --- perhaps the equatorial
density enhancement with lower densities at the poles as favored for a
number of reasons by previous authors (see Monnier et al.\ 1999, Smith
et al.\ 2001, and references therein) --- then it may also help
account for the apparent axisymmetry inferred from molecular studies
with lower spatial resolution as well as requirements from maser
studies.

On the other hand, faster (40--60 km s$^{-1}$) dense cloudlets moving
through this halo are seen more easily at shorter wavelengths in
reflected starlight and at high spatial resolution.  Muller et al.\
(2007) envisioned a schematic geometry with three main components: a
dense compact dusty component, a diffuse extended envelope, and a
bipolar high-velocity component.  Our spectra reveal, however, that
the high-velocity component is made of cloudlets that are intermixed
randomly with the halo, with no clear large-scale axis of symmetry.

The configuration in Figure~\ref{fig:cartoon} is such that the fast
and dense cloudlets dominate images and spectra with sufficient
resolution to see them.  Denser condensations obviously emit and
reflect more light than their less dense surroundings.  Moving more
quickly than their surroundings, they may leave relatively low-density
and lower-extinction wakes that further enhance their illumination by
the star.  A favorably-positioned observer looking through one of the
larger cavities would have a more direct view of the star than we do.
This geometric structure in Figure~\ref{fig:cartoon} would also be
favorable for the production of strong K~{\sc i} emission lines
following the mechanism outlined in the Appendix of Humphreys et al.\
(2005).  It would favor the escape of net emission primarily on the
near side of the nebula, since K~{\sc i} photons traversing from the
far side would likely be resonantly scattered away before making it
through.  In this scenario, then, the K~{\sc i} emission arises
primarily from the surface of last scattering on the boundaries of
dusty molecular cloudlets.  This could explain why the K~{\sc i} and
4.6~$\micron$ CO emission appear to be spatially and kinematically
coincident in the nebula around VY~CMa.  They arise in very different
regions in the smoother stellar wind of Betelgeuse, where calculations
by Huggins et al.\ (1994) predict a clear stratification of atomic and
molecular gas.

%%%%%%%%%%%%%%%%%%%%%%%%%%%%%%%%%%%%%%%%%%%%%%%%%%%%%%%%%%%%%%%%%%%%%%%%%%%%%%%%%%
\section{SUMMARY AND CONCLUSIONS: WHAT WILL WE SEE IF THESE STARS
  EXPLODE?}

In order to use core-collapse SNe as probes of massive-star evolution
in other galaxies at large distances, we must first be able to
confidently map the various types of SNe to their likely progenitor
stars.  This is especially difficult for those with prodigious CSM
interaction because of the wide diversity in SNe~IIn.  It is therefore
instructive to take well-studied and spatially-resolved examples of
observable circumstellar environments around nearby evolved massive
stars and ask what type of SNe we might expect in the event that they
explode.  One can make at least rough estimates of the level of CSM
interaction and its qualitative appearance when the hypothetical SN
blast wave strikes the observed environment.

\subsection{A Betelgeuse Supernova}  % no fuckin way

For Betelgeuse, our observations of spatially-resolved 4.6~$\micron$
CO emission located at offsets of $\beta\simeq$0$\farcs$5 to 5\arcsec\
(roughly 100--1000 AU) from the central star show an intensity that
falls as $\beta^{-3}$.  Density variations (either large-scale
clumping or other inhomogeneities) are not more than about 20\% of the
average density on size scales of $\ga$50 AU.  This indicates an
environment shaped by a predominantly steady wind, with a constant
outflow speed of $V_{CSM}\simeq$15 km s$^{-1}$ and mass-loss rate of
roughly $\dot{M}\simeq$2$\times$10$^{-6}$ $M_{\odot}$ yr$^{-1}$ over
at least the past $\sim$300 yr.

If Betelgeuse were to explode in its current RSG phase, it would
produce a Type II SN.\footnote{However, its luminosity suggests an
  initial mass of 15--20 $M_{\odot}$, much like that of the progenitor
  of SN~1987A. Thus, Betelgeuse could enter a blue loop and explode as
  a BSG after $\sim$10$^4$ yr.}  If the blast wave from that SN
expands at $\sim$15,000 km s$^{-1}$, then it will reach a radius of
10$^{3}$~AU (5\arcsec) in 116~days.  Thus, during the bright plateau
phase of a SN~II-P, for example, the shock will sweep directly through
the portion of the CSM that is resolved in our 4.6~$\micron$ CO data.

All massive stars with winds will give rise to SNe with {\it some}
level of CSM interaction, but would Betelgeuse's CSM interaction be
strong enough to detect in typical SN observations?  The wind density
parameter $w$=$\dot{M}$/$V_{CSM}$ for Betelgeuse is roughly
8.5$\times$10$^{13}$ g cm$^{-1}$.  This is more than two orders of
magnitude below typical estimates of $w$ for SNe~IIn (e.g., Chugai \&
Danziger 1994) and four orders of magnitude below overluminous SNe~IIn
like 2006tf (Smith et al.\ 2008).  The total CSM mass contained within
5\arcsec\ of Betelgeuse (out to radii of 1000 AU, lost in the past
$\sim$320 yr) is only about 10$^{-3}$ $M_{\odot}$.  This is vastly
lower than the expected mass of SN ejecta (probably well over
1~$M_{\odot}$), so all the CSM we detect around Betelgeuse does not
have enough inertia to substantially decelerate the blast wave or tap
into its kinetic energy.  Hence, the fast shock is not likely to be
radiative during the main peak of the light curve, so the CSM
interaction will not subtantially enhance the visual continuum
luminosity and cannot give rise to substantial narrow H$\alpha$
emission from the post-shock gas.  In other words, it would not be
classified as a SN~IIn and would appear instead as a normal SN~II-P or
II-L.

What about emission from the photoionized pre-shock CSM?  The average
density in Betelgeuse's wind at a projected radius of 2\arcsec\ is
$n_H\simeq$1125 cm$^{-3}$.  Even if all the pre-shock CSM mass within
1000 AU were photoionized by the SN, the resulting H$\alpha$
luminosity would be only about 0.4~L$_{\odot}$.  In order to be
clearly detected in the spectrum during the main peak of a SN~IIn,
narrow H$\alpha$ lines from the photoionized CSM typically have
L$_{H\alpha}\gtrsim$10$^5$~L$_{\odot}$ (e.g., Salamanca et al.\ 1998,
2002).

In the unique case where Betelgeuse itself were to explode, it would
be close enough that the CSM interaction region (a few arcseconds)
could actually be spatially resolved from the SN photosphere and might
therefore be observable as an extended X-ray and optical emission-line
source.  If a star like Betelgeuse exploded at typical distances of
nearby SNe (more than 10 Mpc), however, the CSM interaction would be
much harder to detect with typical data quality because it would be
drowned in the photospheric light of the SN.

These arguments above apply to visual-wavelength radiation during the
first $\sim$100 days of the light curve when the SN is bright, because
that's when a SN would be classified as a Type II-P, II-L, or IIn.
Detecting signs of CSM interaction at later times and other
wavelengths is a different story.  After the underlying SN~II-P
photosphere fades, signs of late-time CSM interaction may be visible
in emission (Chevalier \& Fransson 1994) or absorption (Chugai et al.\
2007).  CSM interaction from a Betelgeuse-like SN might be detectable
in X-rays and radio emission.  For example, Pooley et al.\ (2002)
measured X-rays from SN~1999em that led them to infer a progenitor
mass-loss rate of 1--2$\times$10$^{-6}$ $M_{\odot}$~yr$^{-1}$.  This
is similar to Betelgeuse's mass-loss rate, so SN~1999em being a
prototypical SN~II-P reinforces our conclusions above that Betelgeuse
cannot produce a SN~IIn.  Progenitor mass-loss rates of 10$^{-6}$ to
10$^{-5}$ $M_{\odot}$ yr$^{-1}$ are typical for normal SNe~II-P
(Chevalier et al.\ 2006).

\subsection{A Supernova from VY~CMa}

The case of VY~CMa is different, because its current mass-loss rate is
100 times higher than that of Betelgeuse, and its CSM is very
inhomogeneous.  Our 4.6~$\micron$ spectra reveal a locally
inhomogeneous but globally isotropic distribution of CO-emitting gas,
resulting from a series of episodic mass ejections over the past 1000
years that have built up a dense shell environment like that depicted
in Figure~\ref{fig:cartoon}.  This sketch of the CSM is complicated,
but is needed to accomodate the seemingly discrepant observations as
noted earlier (see also Humphreys et al.\ 2005, 2007; Smith 2004;
Smith et al.\ 2001; Monnier et al.\ 1999, 2004). Most of the visible
mass is contained in cloudlets that are either ejected bullets or
swept up arcs and shells, or both.  VY~CMa's current average mass-loss
rate has been estimated as 2--4$\times$10$^{-4}$ $M_{\odot}$ yr$^{-1}$
(Danchi et al.\ 1994), which we attribute to the smoother and slower
wind between clumps.  Our 4.6~$\micron$ CO data and the intensity of
scattered starlight (Smith et al.\ 2001) suggest that between a few
hundred and 1000 years ago, VY~CMa had a higher mass-loss rate of
1--2$\times$10$^{-3}$ $M_{\odot}$ yr$^{-1}$ attributed mainly to the
mass in dense clouds.

With these parameters, the total mass ejected by VY~CMa in the past
$\sim$1000 yr is a little over 1~$M_{\odot}$.  Unlike Betelgeuse, this
may be enough to substantially decelerate the forward shock (depending
on the mass and density law in the SN ejecta), and thereby tap into
the available reservoir of its kinetic energy to power a display of
CSM interaction.

There are basically two components in VY~CMa's wind: If we assume that
(1) the slow ``halo'' expanding at 25 km s$^{-1}$ corresponds to the
current average mass-loss rate of 3$\times$10$^{-4}$ $M_{\odot}$
yr$^{-1}$, and that (2) dense cloudlets moving at 35 km s$^{-1}$
produced the higher mass-loss rate of 2$\times$10$^{-3}$ $M_{\odot}$
yr$^{-1}$, then the wind density parameter $w$=$\dot{M}$/$V_{CSM}$ for
VY~CMa is roughly (1--3)$\times$10$^{16}$ g cm$^{-1}$ (bear in mind
that there may be localized clumps with significantly higher density
than the average).  If most of the mass is in the clumpy wind, then a
value near 2$\times$10$^{16}$ g cm$^{-1}$ is conservative.

Unlike Betelgeuse, VY~CMa's value of $w$ is entirely consistent with
typical values inferred for moderate-luminosity SNe~IIn.  The maximum
luminosity that can be generated by CSM interaction ($L_{CSM}$),
assuming 100\% efficiency in converting kinetic energy into
visual-wavelength radiation, is given by

\begin{equation}
L_{CSM} = \frac{1}{2} w V_{SN}^3 
\end{equation}

\noindent where $V_{SN}$ is the speed of the forward shock sweeping
through the CSM.  Adopting $V_{SN}$=5,000 km~s$^{-1}$ for the
decelerated forward shock speed, CSM interaction in a VY~CMa-like SN
could yield a luminosity potentially as high as 3$\times$10$^8$
L$_{\odot}$ at early times when the conversion of kinetic energy to
light is efficient. This could at most double the peak continuum
luminosity of a normal SN~II-P at early times, but would likely
dominate the appearance of the spectrum with strong and relatively
narrow H$\alpha$ emission.  Thus, an exploded VY~CMa would likely
appear to a distant observer as a {\it bona-fide} SN~IIn.  The ongoing
CSM interaction would also dominate the luminosity at late times
(after $\sim$1 yr).  With $V_{SN}$=5,000 km~s$^{-1}$, the
$\sim$1~$M_{\odot}$ residing within 7500 AU of the star can be
overtaken by the forward shock by 6--8 years after the SN.  Thus,
while VY~CMa is unlikely to produce an extremely luminous SN~IIn like
SN~2006tf (Smith et al.\ 2008), it probably will produce a moderately
luminous and long-lasting SN~IIn like SN~1988Z, which had detectable
CSM interaction for over a decade after explosion (Turatto et al.\
1993; Chugai \& Danziger 1994; Aretxaga et al.\ 1999).

In fact, both the qualitative depiction in Figure~\ref{fig:cartoon}
and our quantative estimates for VY~CMa's CSM provide a surprisingly
good match to the CSM that Chugai \& Danziger (1994) envisioned for
SN~1988Z's progenitor.  Chugai \& Danziger (1994) interpreted SN~1988Z
as having a lower-mass (8--10~$M_{\odot}$) progenitor star than
VY~CMa, but the inferred values of $w$=5$\times$10$^{16}$ g~cm$^{-1}$
and $\dot{M}$=7$\times$10$^{-4}$ $M_{\odot}$~yr$^{-1}$ for the clumpy
progenitor wind are very close to those of VY~CMa.  In particular, the
two-component clumpy wind required to enhance the emissivity of
intermediate-width post-shock H$\alpha$ emission lines of SN~1988Z are
realized in the circumstellar nebula of VY~CMa
(Fig.~\ref{fig:cartoon}).

This enhanced episodic mass loss by VY~CMa apparently just turned on
about 1000 yr ago, judging by the outer extent of its visible
reflection nebula (Smith et al.\ 2001) and the steep fall off of
$\beta^{-6}$ at 5\arcsec\ (7500 AU) for 4.6~$\micron$ CO emission
(Fig.~\ref{fig:radVY}).  Thus, after a few years we might expect the
enhanced luminosity from CSM interaction to shut off when the blast
wave overruns the outer boundary of the nebula.  This too is
reminiscent of some SNe~IIn.  For example, a sharp drop in visual
luminosity was attributed to the forward shock overtaking a similar
outer boundary of a dense shell in the case of the SN~IIn 1994W
(Chugai et al.\ 1994), while a sharp change in pre-SN mass loss has
also been inferred for SNe~1980K and 1988Z from an analogous drop in
radio emission (Montes et al.\ 1998; Williams et al.\ 2002).

As an extreme RSG, VY~CMa is valuable in that it provides a
well-studied example that LBVs are not necessarily the {\it only}
candidates for the progenitors of SNe~IIn, which has been suggested
recently (see \S 1).  An important point, though, is that VY~CMa's
unusually dense CSM is not the product of a steady RSG wind, but
instead, resulted from an episode of enhanced eruptive mass ejection
in just the past 1000 yr.  Thus, whether LBVs or extreme VY~CMa-like
RSGs are the progenitors of SNe~IIn, {\it eruptive} rather than {\it
  steady} pre-SN mass loss is apparently an essential ingredient for
sufficient CSM interaction.  This may be a critical clue for
understanding the final stages of evolution for massive stars.

This only applies if VY~CMa will explode soon.  There is, of course,
considerable uncertainty as to the true fate of VY~CMa, since the late
evolution of massive stars is still poorly understood and depends
sensitively on the mass loss.  Will VY~CMa explode in its current
extreme RSG state, or will it evolve blueward to become a yellow
hypergiant with a dense environment (like IRC+10420), or a blue
supergiant with a dense shell nebula in a post-RSG LBV phase (Smith
2007; Smith et al.\ 2004)?  Or instead, will it voyage all the way to
the WR phase, producing a SN~Ib with a distant wind-swept shell?

VY CMa is admittedly a rare and extreme example, even among the most
luminous RSGs, and has the most prodigious mass loss among stars with
initial masses 20--40~$M_{\odot}$.  We can infer one of two
possibilities: Either (1) only a small fraction of SNe from stars of
initial mass 30--40~$M_{\odot}$ will ever have dense enough CSM to
yield SNe~IIn, or (2) since VY~CMa's unusually high mass loss just
``turned on'' about 1000 years ago, it may be symptomatic of the brief
enhanced mass loss inferred for pre-explosion evolution of many
SNe~IIn (e.g., Chugai et al.\ 1994; Smith et al.\ 2008).  In the
latter case --- where we really are witnessing the ``last gasps'' of
VY~CMa (Monnier et al.\ 1999) --- G.\ Wallerstein's gamble in the
1960s to continue monitoring VY~CMa until it explodes (see Wallerstein
\& Gonzalez 2001) may yet pay off.

\acknowledgments \scriptsize

We thank Graham Harper for interesting discussions about the
Betelgeuse data and interpretations, and for helpful comments on the
manuscript.  We also thank Phil Hinz for permission to use the IR
image of Betelgeuse that appears in Figure 1.  This work was based on
observations obtained at the Gemini Observatory, which is operated by
the Association of Universities for Research in Astronomy, Inc., under
a cooperative agreement with the NSF on behalf of the Gemini
partnership: the National Science Foundation (US), the Particle
Physics and Astronomy Research Council (UK), the National Research
Council (Canada), CONICYT (Chile), the Australian Research Council
(Australia), CNPq (Brazil), and CONICET (Argentina).  The observations
were obtained with the Phoenix infrared spectrograph, which was
developed and operated by the National Optical Astronomy Observatory.
N.R.\ is a Royal Swedish Academy of Sciences research fellow supported
by a grant from the Knut and Alice Wallenberg foundation. The spectra
were obtained as part of programs GS-2005A-C-8 (P.I.: Harper) and
GS-2006B-DD-1 (P.I.: Hinkle).

% REFERENCES

\end{document}